\documentclass[
    a4paper,
    10pt,
    twocolumn
]{article}

\usepackage[
    a4paper,
    margin=2cm
]{geometry}

\usepackage[utf8]{inputenc}
\usepackage[T1]{fontenc}
\usepackage[english]{babel}
\usepackage{mathptmx}

\usepackage{amsmath,amssymb}
\usepackage{bm}
\usepackage{dcolumn}
\usepackage{siunitx}

\usepackage{graphicx}
\usepackage{booktabs}
\usepackage{xcolor}
\usepackage{tikz}
\usepackage{placeins}

\usepackage[caption=false]{subfig}

\usepackage{authblk}

\usepackage{etoolbox}
\usepackage{nomencl}

\makenomenclature

\renewcommand{\nomgroup}[1]{%
  \item[\bfseries
  \ifstrequal{#1}{A}{Acronyms}{%
  \ifstrequal{#1}{S}{Symbols}{%
  \ifstrequal{#1}{U}{Subscripts}{%
  \ifstrequal{#1}{P}{Superscripts}{}}}}%
  ]}

\usepackage[authoryear,round]{natbib}

\usepackage{lineno}

\usepackage{hyperref}

\title{Impact of a resonator on vortex induced vibrations
	of a wind turbine airfoil}

\author[1]{S. G. Horcas%
	\thanks{Corresponding author:
		\texttt{sergio.gonzalez-horcas@upc.edu}}}

\author[1,2]{D. Roca}

\author[1,3]{E. Ortega}

\author[1,2]{J. Cante}

\affil[1]{%
	Department of Physics (FIS),
	Universitat Polit\`ecnica de Catalunya -- BarcelonaTech (UPC),
	Colom 1--11, 08222 Terrassa, Spain
}

\affil[2]{%
	Computational Design \& Analysis of Engineering Metamaterials Group,
	International Centre for Numerical Methods in Engineering (CIMNE),
	Gran Capit\`a s/n, 08034 Barcelona, Spain
}

\affil[3]{%
	Aerospace Engineering Group,
	International Centre for Numerical Methods in Engineering (CIMNE),
	Gran Capit\`a s/n, 08034 Barcelona, Spain
}

\begin{document}

\maketitle

\begin{abstract}

	The aeroelastic instability known as vortex induced vibrations (VIV) has its origin in the coupling of bluff body shedding and the structure dynamics, and can have adverse effects on several industrial fields. In this work the use of locally resonant metamaterials (LRM) is explored as a wind turbine blade VIV mitigation strategy. The problem is cast into a classical airfoil in transverse oscillation, elastically mounted. A fluid structure interaction model is then generated and validated for the main system. This is achieved using an open source finite volume code with arbitrary Lagrangian Eulerian capabilities and a built-in rigid body motion solver. Subsequently, the capabilities of the code are extended by coupling the dynamics of a single resonator into the main system through a tailored development. The performance of the resonator was evaluated for different designs, which were parameterized mainly by the resonator frequency and its relative mass with respect to the airfoil. Analysis of the results identified a frequency band where the resonator effectively influenced vibration amplitudes. However, the amplitude attenuation was very low when considering realistic relative resonator masses. In particular, achieving a 50\% amplitude reduction required a mass ratio larger than 0.25. Studying the mechanisms uncovered the cause of these beneficial effects, which relied on the aeroelastic frequency bifurcation introduced by the resonator. These fundamental findings open the door to future applications of alternative technologies for VIV suppression.

\end{abstract}

\nomenclature[A]{CFD}{Computational fluid dynamics}
\nomenclature[A]{LRM}{Locally resonant metamaterials}
\nomenclature[A]{VIV}{Vortex induced vibrations}
\nomenclature[A]{LCO}{Limit cycle oscillations}
\nomenclature[A]{STD}{Standard deviation}
\nomenclature[A]{AoA}{Angle of attack}
\nomenclature[A]{RANS}{Reynolds averaged Navier Stokes}

\nomenclature[S]{$d$}{Damping coefficient}
\nomenclature[S]{$\xi$}{Damping ratio}
\nomenclature[S]{$y^+$}{Non dimensional wall distance}
\nomenclature[S]{$F$}{Aerodynamic force}
\nomenclature[S]{$k$}{Stiffness}
\nomenclature[S]{$m$}{Mass}
\nomenclature[S]{$c$}{Chord}
\nomenclature[S]{$t$}{Time}
\nomenclature[S]{$x$}{Position}
\nomenclature[S]{$U_\infty$}{Freestream velocity}
\nomenclature[S]{$f$}{Frequency}
\nomenclature[S]{$\omega$}{Angular frequency}
\nomenclature[S]{$C_l$}{Lift coefficient}
\nomenclature[S]{$C_d$}{Drag coefficient}
\nomenclature[S]{$C_p$}{Pressure coefficient}
\nomenclature[S]{$\rho$}{Density}
\nomenclature[S]{$\gamma$}{Ratio resonator and airfoil}
\nomenclature[S]{$U^*$}{Reduced velocity, $U^*=U_\infty/(c f)$}
\nomenclature[S]{$f^*$}{Reduced frequency, $f^*=1/U^*$}
\nomenclature[S]{$A$}{Amplitude of main frequency}
\nomenclature[S]{$A^*$}{Normalized amplitude, $A^*=A/c$}
\nomenclature[S]{$\nu$}{Kinematic viscosity}

\nomenclature[U]{$a$}{Airfoil or main system}
\nomenclature[U]{$r$}{Resonator}
\nomenclature[U]{$sys$}{Coupled airfoil and resonator system}
\nomenclature[U]{$natsh$}{Natural shedding}

\printnomenclature

\section{Introduction}
Some slender large structures, when subjected to particular aerodynamic conditions, are prone to accentuated aeroelastic effects. This significant aeroelastic response implies not only a potential increase in static deflections, but also a higher susceptibility to aeroelastic instabilities. In this category we find the so-called vortex induced vibrations (VIV), typically studied for bridge decks and cables [\cite{bridgecable,bridgedeck}] and which have been identified as one of the key challenges limiting the progress of large horizontal axis wind turbines [\cite{veers_grand_2022}]. VIV is a flow induced vibration phenomenon that involves massively separated flow and a strong interaction between the flow shedding and structural frequencies. In the context of wind energy, it is particularly concerning for the tower and the rotor blades [\cite{Horcas2022POF, Horcastipgeom}], where the first edgewise mode can be heavily excited in standstill conditions (i.e. in the absence of rotation).
Previous numerical simulations of blades VIV have predicted large, potentially sustained edgewise vibration amplitudes that compromise the structural integrity of the system. Furthermore, mitigation strategies based on control surfaces seem to have limited effectiveness [\cite{Horcasflap}]. Therefore, the development of alternative and efficient suppression strategies is of primary importance. In this work, the use of locally resonant metamaterials (LRM) for wind energy VIV mitigation is explored for the first time. Such metamaterials hold the potential to attenuate the dynamics of certain frequency bands by introducing exotic properties to a given baseline material [\cite{metamat}].
Consequently, they are foreseen as potential lightweight solutions for various applications.
Perhaps the most studied application in this regard is the acoustic insulation, where \cite{roca_computational_2019} conducted numerical studies where flat panels were designed with the help of topological optimization. An overview of the potential of acoustic metamaterials for cabin noise mitigation is also provided in~\cite{ang_acoustic_2016}. Metamaterials in the form of Phononic Subsurfaces (PSubs) have also been proposed in the context of flow control for stabilizing boundary layer perturbations [\cite{hussein2015}]. The application of LRM to flow induced vibrations is more scarce, but it includes the experimental work of~\cite{pires_use_2022} for the attenuation of a flat plate subjected to the vortex shedding of an upstream cylinder. LRM have also been used in the context of fluid structure interaction in attached flow [\cite{casadei_harnessing_2014, casadei_wave_2014}], relying on airfoil-type units and performing both numerical and experimental analyses. Finally, they have been studied for the vibration of rotating beams [\cite{basta_vibration_2020}], where even relatively lightweight resonators achieved significant vibration attenuation.

Traditionally, local resonators are modeled through a series of spring-mass subsystems, eventually tuned by parametric variations or through numerical optimization (see e.g. \cite{xiao_energy_2023,casalotti_metamaterial_2018,das_bendingtorsion_2023}). In practice, the intended behavior of local resonators is realized through the introduction of periodic microstructures on the surface [\cite{zhu_experimental_2011}], or by embedding them in a lattice structure, such as a chiral topology [\cite{zhu_chiral_2014,liu_wave_2011}]. The advantage of the latter approach is that the metamaterial is included within the structure, thus remaining transparent to the aerodynamic behavior. This differs from other approaches for VIV mitigation, such as the use of helical strakes [\cite{ZHOU2011903}] or the more recent research track concerning metasurfaces by \cite{wang_use_2021}. These potential benefits of vibration attenuation while remaining aerodynamically transparent and passive (contrary to complex active solutions deployed e.g. for flutter in \cite{activeflutter}) motivated the exploratory study performed in this work.

\begin{figure*}[!htbp]
	\centering
	\includegraphics[width=0.7\textwidth]{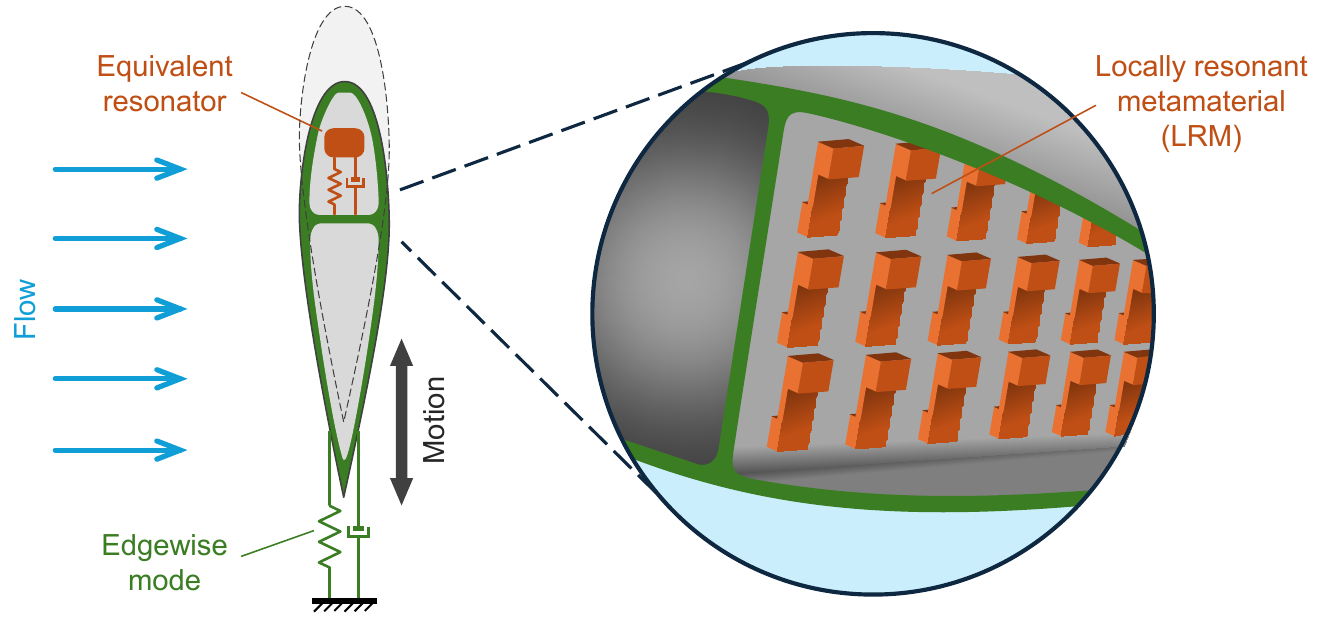}
	\caption{Diagram of studied aeroelastic system.}
	\label{f:object}
\end{figure*}

With the aim of providing a first estimation of the impact of metamaterials on vibrations, they were modeled in this work as simple resonators attached to a structure already undergoing VIV
Despite its simplification, this modelling approach is commonly employed when exploring the effects of local resonance phenomena attributed to metamaterials. For instance, in \citet{harris2026} a PSub for broadband flow control is modelled by means of spring-mass devices which are then linked to a final realistic structure through inverse design.
The reference VIV problem was simplified from a wind turbine blade by casting it into the classical problem of a bluff body in transverse oscillation (Figure~\ref{f:object}).
In this setup, the main elastic mounting mimics the edgewise mode of a wind turbine blade, which has been shown to dominate over flapwise and torsional modes in terms of energy transfer [\cite{Horcas2022POF}].
An airfoil was chosen to be representative of a blade section. This also represents a considerable simplification since three dimensional effects are neglected, but allowed for a first efficient exploration of the main expected aeroelastic mechanisms. The resonators were then added as a coupled subsystem, simplified in a first approach as another spring and damper with its own parameters.
In this way, the studied configuration resembles the tuned mass damper approach installed for heavy circular cylinders in \cite{LUO2022103692}, where promising vibration attenuations were observed even for relative device to main system masses of 10\%.

To the best of the authors knowledge, the present work represents the first exploration of metamaterials for direct exploitation in vortex induced vibrations of airfoils. Even at this early stage of the application analysis, it was decided to model the flow by means of a state of the art computational fluid dynamics (CFD) finite volume solver. The equations of motion representing the coupling of these nonlinear aerodynamic loading with the elastic mounting were solved during run time. This was done adopting a fluid structure interaction (FSI) approach, which accounted for the modeling of the resonators and their coupling with the main system. This approach also allowed for the validation of the baseline experiment (i.e. before the introduction of the resonators).
The manuscript addresses the particular aeroelastic mechanisms exerted by the metamaterial, and sheds light on the potential of such technology from an aeroelastic point of view. Particular attention was paid to the characteristic frequencies of the problem (structural, aerodynamic, and aeroelastic). Their complex interaction was analyzed in view of the response of the baseline airfoil, as well as with the subsequent addition of resonators.

\section{Methodology}

Section~\ref{ch:aerosolver} introduces the solver used for the aerodynamics, together with the main modeling choices made in terms of the flow. This description is complemented by Section~\ref{ch:fsi}, where details about the aeroelastic coupling are introduced. This includes the development performed within the framework of this study, which allowed for the modeling of the resonator. Finally, Section~\ref{ch:themodel} describes the grid employed and the different configurations chosen to be studied, together with the modeling parameters. Since the necessary parameters and test results were available, the baseline configuration (without metamaterials) aimed to reproduce the vortex induced vibrations from the symmetric NACA 0021 experiments of \cite{BENNER2019102577}.

\subsection{Flow solver}
\label{ch:aerosolver}

The open source CFD package \texttt{OpenFoam} [\cite{OpenFOAM11}] was used to solve the incompressible Navier Stokes equations on unstructured grids. A finite volume approach was employed, assuming a Newtonian fluid and neglecting gravity loads. For simulations accounting for aeroelastic coupling, the arbitrary Lagrangian Eulerian (ALE) formulation was adopted. Turbulence modeling relied on an unsteady Reynolds-Averaged Navier Stokes (URANS) approach, starting from steady simulations. Steady runs were solved \textit{via} the semi implicit method for pressure linked equations (SIMPLE), while the unsteady counterparts used the pressure implicit with splitting of operators (PISO) algorithm. To maximize the computational efficiency of the time marching simulations, a variable time step approach was adopted, steered by a maximum Courant number. The $k$-$\omega$ SST turbulence model was employed without a wall model. Therefore, a value of $y^+<1$ was targeted during mesh generation. Numerical schemes were chosen to be second order accurate in time and space.

\subsection{Fluid structure interaction}
\label{ch:fsi}

\begin{figure}[!htbp]
	\centering
	\subfloat[Main airfoil (no resonator setup, \texttt{No r})]{\includegraphics[width=0.25\textwidth]{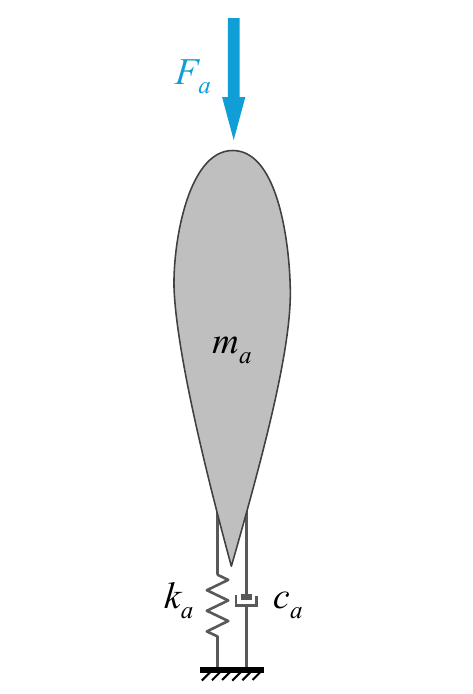}\label{f:setup1}}
	\qquad
	\subfloat[Main airfoil + resonator]{\includegraphics[width=0.25\textwidth]{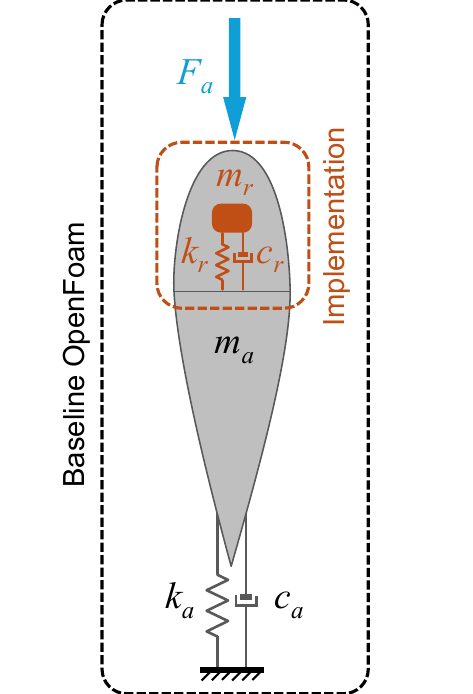}
		\label{f:setup2}}
	\caption{Aeroelastic setups considered for this work}
	\label{f:setup}
\end{figure}

Two main aeroelastic setups were considered in the present work, as presented in Figure~\ref{f:setup}. These configurations represent a simplification of more complex setups, such as a mode shape. One setup features the airfoil mounted on a simple spring damper system. The other is coupled with the resonator, allowing for the assessment of its effect on vibrations. For both configurations, the elastic mounting and the CFD solver exchanged displacements and loads, respectively, in a staggered and loosely coupled fashion. The system motion was understood as a wall displacement by the CFD code, and applied as a Dirichlet boundary condition for a mesh deformation problem solved in \texttt{OpenFOAM}. The nodes of the external edge of the CFD domain were assumed to be fixed for this mesh deformation process, and spherical linear interpolation (SLERP) was employed. The aeroelastic problem was solved with a symplectic second order explicit time integrator.
The configuration accounting only for the main system, shown in Figure~\ref{f:setup1}, was completely modeled by means of the built-in \texttt{OpenFOAM} features.
In particular, a six degrees of freedom motion model was employed, though restricted in this study solely to the vertical direction.
This model assumes a rigid body and allows for the attachment of an elastic mounting. It accounted for an elastic system (i.e. a translational spring $k_a$ and damper $d_a$) attached to the \texttt{Airfoil} mass $m_a$. This system was subjected to CFD loads $F_a$ in the direction of the axial spring, which in this particular case corresponded to the lift. To model the effects of the resonator, the main system was extended as depicted in Figure~\ref{f:setup2}. This extension consisted of the introduction of a spring $k_r$ and a damper $d_r$, connecting $m_a$ to the mass of the resonator $m_r$. As a simplification, external loads on $m_r$ were disregarded.

The interaction of the resonator with the main system required a tailored modification of the \texttt{OpenFOAM} baseline capabilities. This relied on a second order finite differences approach with a variable time step, which tracked the displacement of the resonator $x_r$ and injected the corresponding loads into the \texttt{Airfoil} system. This process was performed once per time step following a loosely coupled approach, resulting in the system summarized in Equation~\ref{e:aerosystem}.

\begin{equation}
	\begin{aligned}
		 & m_a \ddot{x}_a + d_a \dot{x}_a + k_a x_a - d_r \left ( \dot{x}_r - \dot{x}_a \right ) - k_r (x_r - x_a) = F_a \\
		 & m_r \ddot{x}_r + d_r \left ( \dot{x}_r - \dot{x}_a \right ) + k_r (x_r - x_a) = 0
	\end{aligned}
	\label{e:aerosystem}
\end{equation}
with the damping of each component $i$ being computed by means of the damping factor \textit{via} $d_i = 2 \xi_i \sqrt{ k_i m_i}$. The implementation of this additional feature in \texttt{OpenFOAM} was verified using analytical solutions. These were derived for specific configurations, namely by neglecting $F_a$ and damping. A summary of this study can be found in Appendix~\ref{ch:verification}.

\subsection{Numerical model}
\label{ch:themodel}

\begin{figure}[!htbp]
	\centering
	\fbox{\includegraphics[height=\columnwidth, angle=90]{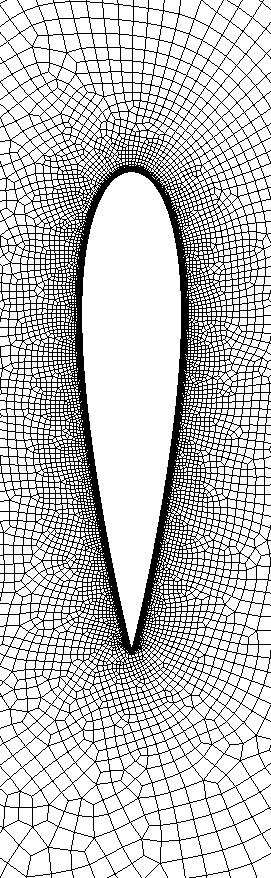}}
	\caption{Mesh zoom}
	\label{f:themesh}
\end{figure}

The grids of the airfoils in this study were parametrically generated using \texttt{gmsh} [\cite{gmsh}]. They had a hybrid topology, with a structured region corresponding to the boundary layer (generated by extrusion with a constant expansion factor of 1.2) and a quad-dominated topology elsewhere. The latter region was generated by means of a frontal Delaunay algorithm for triangles, followed by a subsequent recombination into quads.
The CFD domain was designed to be rectangular, featuring an inlet/outlet strategy for the vertical boundaries. A slip boundary condition was used for the top and bottom boundaries, and a no slip condition was applied to the airfoil surface. The inlet was located at 15$c$ from the \texttt{Airfoil}, and the outlet was at 50$c$. The upper and bottom surfaces were located at 20$c$.
The first cell height was set to $2.5 \times 10^{-4}$ $c$, consistent with the requirements imposed by the lack of wall models. A mesh refinement was carried out in the vicinity of the leading edge and trailing edge. Furthermore, an intermediate region (defined by wall distance) was introduced to avoid expanding the mesh size too suddenly around the airfoil wake. The resulting mesh, as depicted in Figure~\ref{f:themesh}, accounted for approximately \num{30000} elements. Prior to the massive separation simulations for the NACA0021 airfoil, the same methodology was tested for relevant airfoils in wind energy, both in attached flow and in the early stages of stall. Appendix~\ref{ch:valnomounted} summarizes the validation exercises for these unmounted configurations.

For the configurations aiming to study VIV and the effects of the resonator, the airfoil was placed at a 90~\textdegree angle of attack, consistent with the experiments. The parameters used in the tests had to be cast into a non dimensional framework. This involved assuming some details of the experimental setup and casting it into a pure 2D numerical model. In the present work, the runs assumed $c=1.0$, $U_{\infty}=1.0$, $\rho=1.0$, and span $s=1.0$ (SI), and the non dimensional parameters were computed accordingly. The casting of the experimental setup was done on the basis of parameter similarity, which allowed for the selection of $m_a$, $k_a$, and $\xi_a$:

\begin{itemize}
	\item \textit{Stiffness}: The same reduced velocities studied in the experiments were tested in this work. These are defined as $U^*=U_{\infty}/(f_{nw} c)$, where $f_{nw}$ is the structural natural frequency in still water. For a 90\textdegree angle of attack, the experiments reported $f_{nw}=0.59$ Hz. This is very close to the natural frequency in air ($f_{na}=0.61$ Hz), indicating a relatively low added mass effect. Therefore, the CFD runs assumed $f_{nw}=\frac{1}{2\pi}\sqrt{k_a/m_a}$, allowing $k_a$ to be tuned.
	\item \textit{Flow}: Each $U^*$ was related to a freestream Reynolds number $Re=(U_{\infty} c)/\nu$, extracted from the experiments as $Re=(U^* f_{nw} c^2)/\nu$, which was in the range $[1, 12] \times 10^3$. A turbulence intensity of 1\% was assumed, consistent with reported experimental levels.
	\item \textit{Mass}: In the experiments, the mounting and fluid ratio $m^*=m / \left ( \rho V \right )$ was given for the full apparatus with a value of 4.82.
	      The present work assumed that the entire mass was concentrated in the airfoil.
	      This allowed $m_a$ to be computed by estimating the volume $V$ as the area of the NACA 0021 times the span of the test model.
	\item \textit{Damping}: The structural damping ratio $\xi_a$ was chosen to be 0.0067 for the CFD simulations, as it was originally reported in the experiments.
\end{itemize}

The sensitivity of the mesh was also tested for massive flow separation involving VIV prior to reproducing the experimental cases of \cite{BENNER2019102577} (see Appendix~\ref{ch:sensitivities}). Although the results showed some sensitivity to the chosen refinement, the grid was found to be reasonably accurate given the exploratory nature of this study and the high number of parameter variations involved.
A maximum Courant number of 0.9 was used for the unsteady simulations. This corresponded to a time step length of approximately $10^{-4}$~s during limit cycle oscillations (LCO), leading to about $5 \times 10^4$ points per main oscillation period. As shown in the sensitivity study also included in Appendix~\ref{ch:sensitivities}, reducing the maximum Courant number had no noticeable effect on the aeroelastic simulation results. Each final run required approximately 30 hours of wall clock time when running on nine cores of an in-house cluster (Intel Xeon Silver 4214, 2.2-3.2~GHz).

\section{Results}

Before the addition of the resonator, simulations of the baseline case (i.e. the airfoil on the elastic mounting) were performed to characterize the VIV region, as summarized in Section~\ref{ch:baselineres}. This is followed by Section~\ref{ch:resultsresonator}, which covers the introduction of the resonator into the aeroelastic system as an add-on.

\subsection{Airfoil vibrations}
\label{ch:baselineres}

\begin{figure}[!htbp]
	\centering
	\includegraphics[width=\columnwidth]{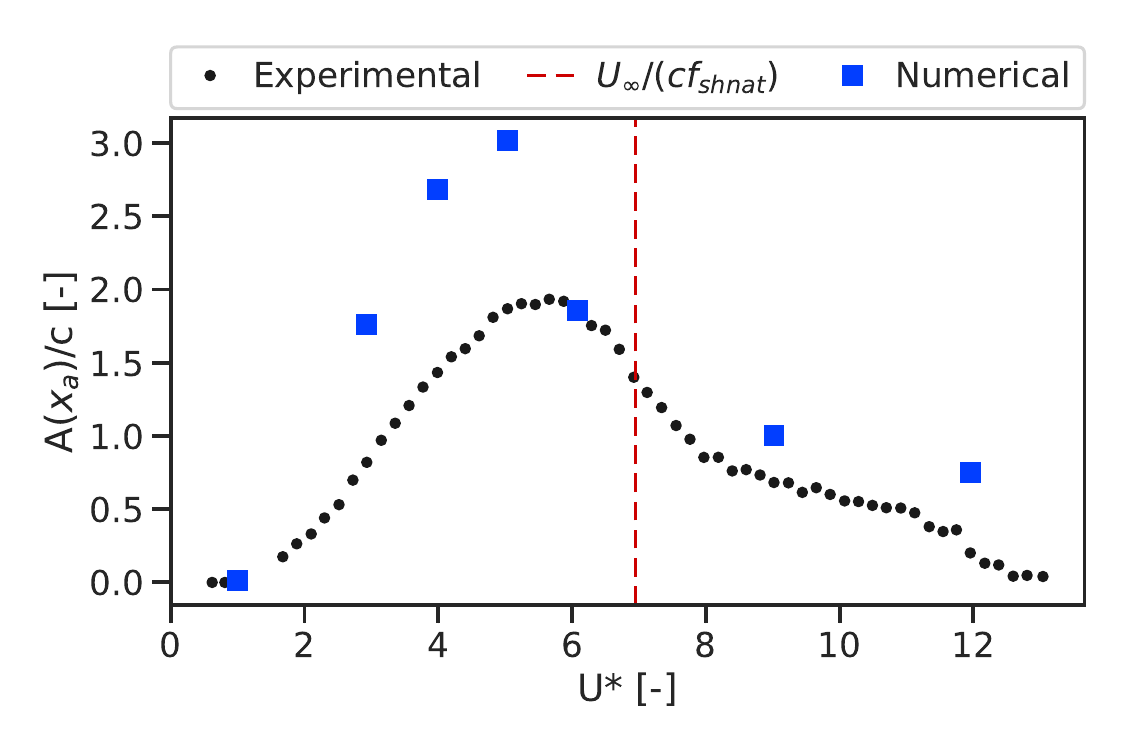}
	\caption{VIV region of the \texttt{Experimental} study of \citet{BENNER2019102577} and of the \texttt{Numerical} results of this work}
	\label{f:baselineres}
\end{figure}

Figure~\ref{f:baselineres} superimposes the relative motion amplitudes $A/c$ from the present study against the experimental results of \citet{BENNER2019102577}. The numerical amplitudes around the main frequency were obtained by means of Lomb-Scargle periodograms. This comparison demonstrates the capability of the present method to accurately capture the amplitude amplification region driven by VIV.
The most notable difference is the large disparity in the maximum vibration amplitude, which was already significantly high in the experiments.
While these differences are left for further exploration as they are not critical to the current study, they are hypothesized to originate from the choices made when translating the experimental setup into numerical model parameters. Specifically, these include concentrating the mass on the airfoil, neglecting the added mass, and assuming a purely 2D domain (whereas no endplates or splitters appeared to be used in the experiments).
For reference, Figure~\ref{f:baselineres} also includes the reduced velocity corresponding to the natural shedding frequency $f_{shnat}$ (i.e. the frequency considering a fixed airfoil). The significant gap between the maximum observed displacements and this frequency reveals a large shift in the lock-in frequency, which is unusual for 2D shapes undergoing typical VIV oscillation.

\begin{figure}[!htbp]
	\centering
	\subfloat[U*=3]{\includegraphics[width=0.32\textwidth]{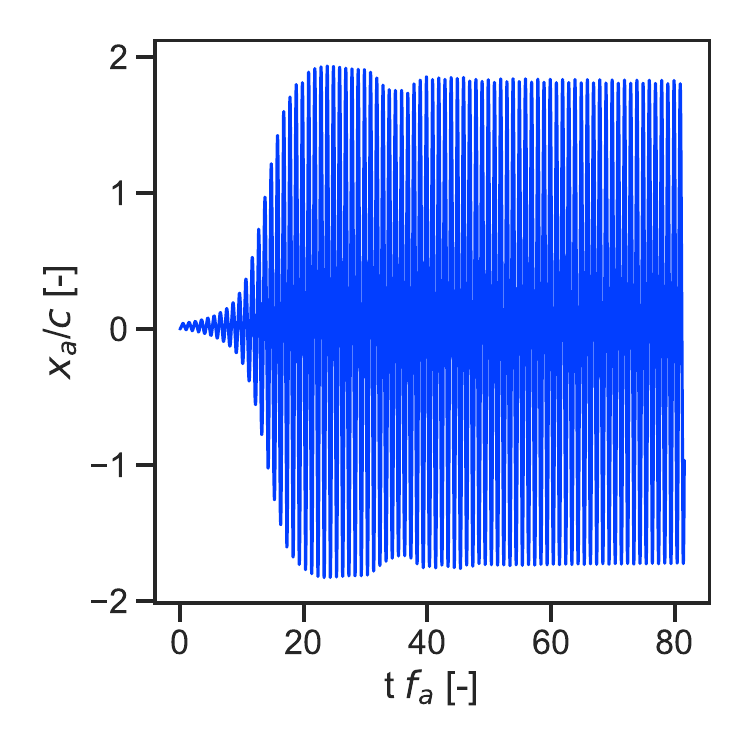}\label{f:baselinerest1}}
	\\
	\subfloat[U*=5]{\includegraphics[width=0.32\textwidth]{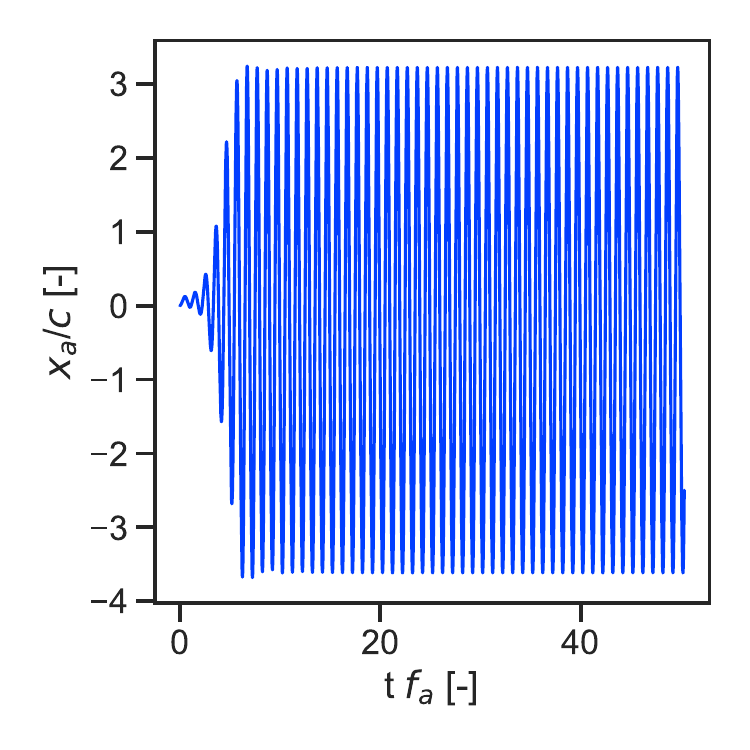}\label{f:baselinerest2}}
	\\
	\subfloat[U*=9]{\includegraphics[width=0.32\textwidth]{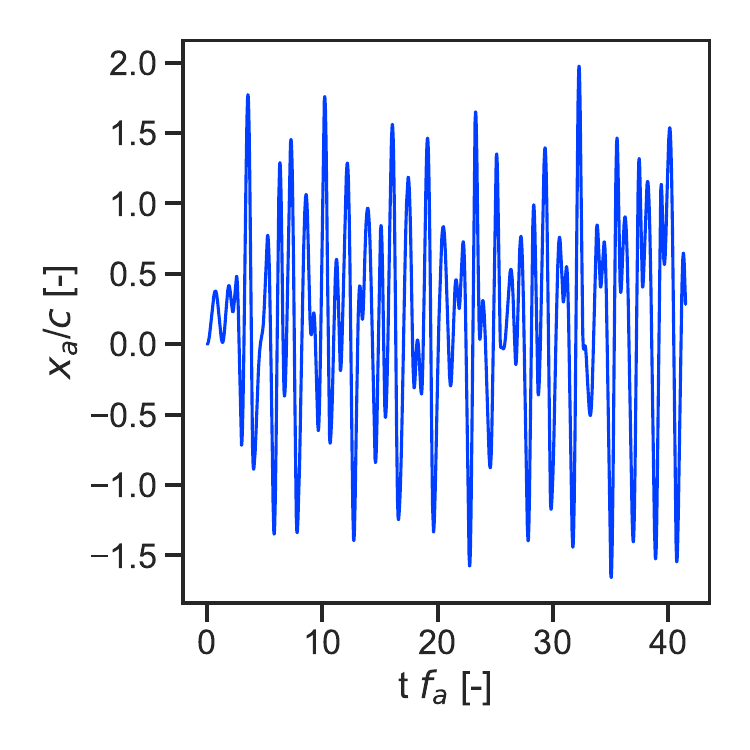}\label{f:baselinerest3}}
	\caption{Airfoil displacement timeseries for selected $U^*$.}
	\label{f:baselinerest}
\end{figure}

\begin{figure}[!htbp]
	\centering
	\subfloat[U*=3]{\includegraphics[width=0.33\textwidth]{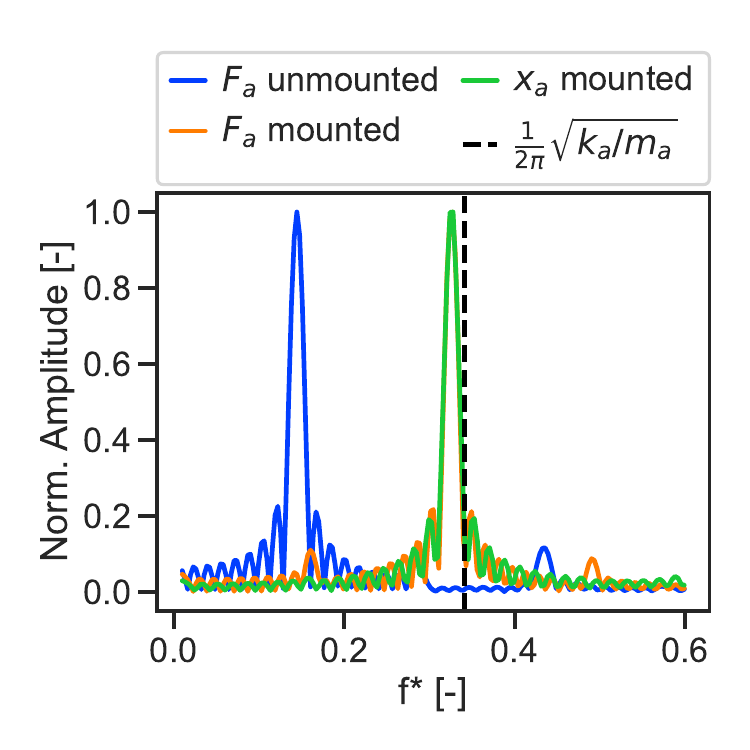}\label{f:baselineresf1}}
	\\
	\subfloat[U*=5]{\includegraphics[width=0.33\textwidth]{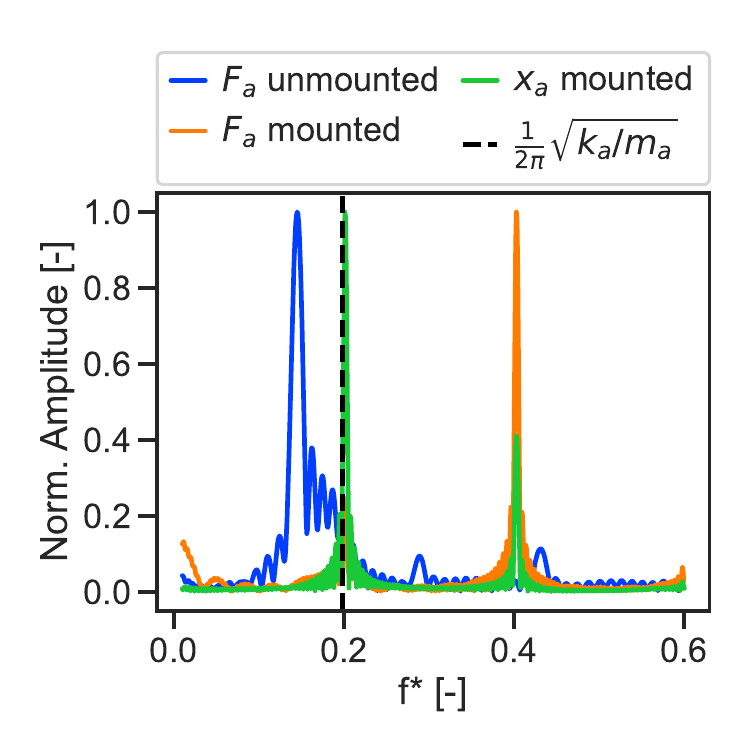}\label{f:baselineresf2}}
	\\
	\subfloat[U*=9]{\includegraphics[width=0.33\textwidth]{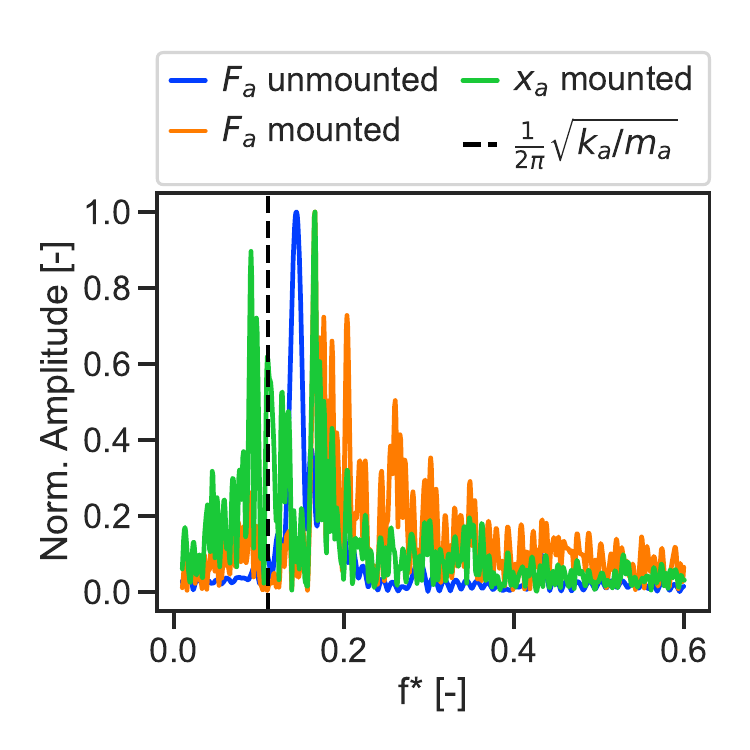}\label{f:baselineresf3}}
	\caption{Normalized characteristic frequencies for selected $U^*$ values. Includes \texttt{unmounted} (i.e. fixed geometry) and elastically \texttt{mounted} configurations. Dashed lines refer to the natural structural frequencies of such mounting.}
	\label{f:baselineresf}
\end{figure}

\begin{figure}[!htbp]
	\centering
	\subfloat[U*=1]{\fbox{\includegraphics[width=0.25\textwidth]{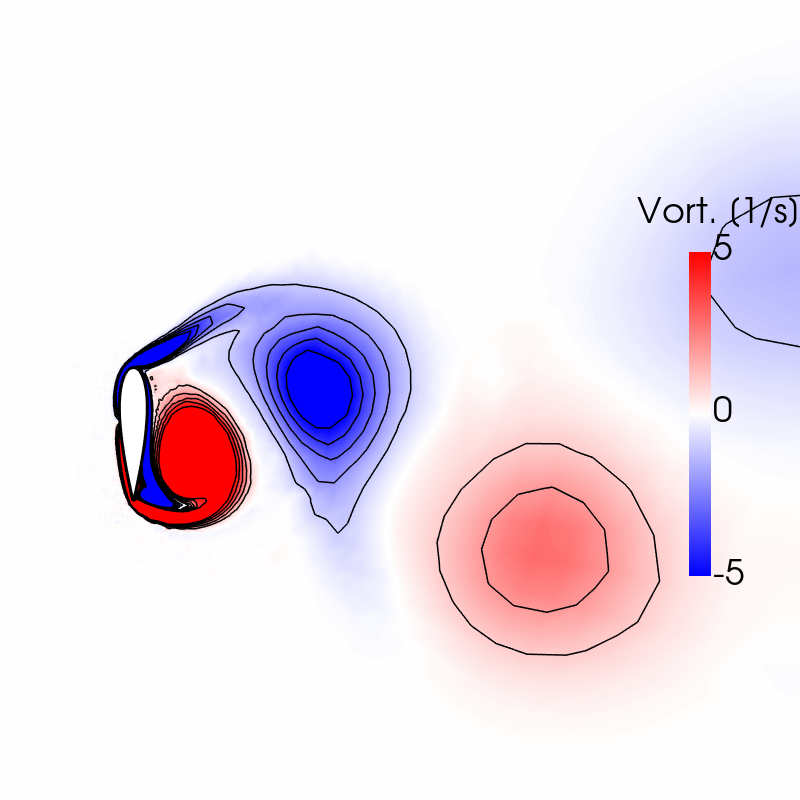}}\label{f:flowviz1}}
	\\
	\subfloat[U*=6]{\fbox{\includegraphics[width=0.25\textwidth]{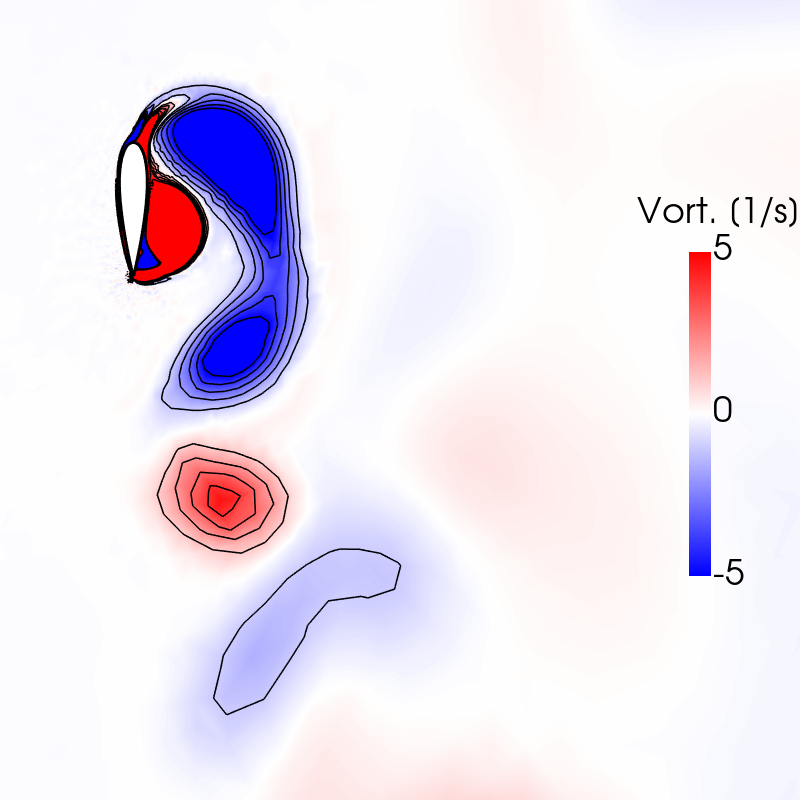}}\label{f:flowviz2}}
	\caption{Instantaneous plot of vorticity for selected vibrating cases of the main system (no resonator). Selected $U^*$ values.}
	\label{f:flowviz}
\end{figure}

Time series of airfoil displacement for selected reduced velocities are presented in Figure~\ref{f:baselinerest}.
The appearance of large amplitudes of motion was related to a sustained vibration characterized by a synchronization of the frequencies of the problem, which is a distinct signature of the VIV phenomenon. The full synchronization of the vibration at $U^*=5$ (Figure~\ref{f:baselinerest2}) was also accompanied by a shorter transient period to reach the limit cycle oscillation (LCO), when compared to other points that also exhibited relatively high motion amplitudes (Figure~\ref{f:baselinerest1}). On the other hand, for points outside the lock-in region, the displacement time series was characterized by multiple frequencies without reaching an LCO (Figure~\ref{f:baselinerest3}).
This frequency behavior is better illustrated through the corresponding frequency spectra for the selected simulations depicted in Figure~\ref{f:baselineresf}. For reference, the figure also includes the spectra for a set of equivalent simulations where the airfoil was not attached to an elastic mounting (labeled as \texttt{unmounted}).
The existence of different characteristic frequencies for the flow and structure is shown in Figure~\ref{f:baselineresf3}.
For the reduced velocities exhibiting lock-in (Figure~\ref{f:baselineresf1} and Figure~\ref{f:baselineresf2}), the flow frequency shifted from the natural shedding frequency to approximately the structural natural frequency (indicated by the vertical black dashed line), in harmony with traditional observations of VIV in transverse motion.
However, contrary to canonical situations, the lock-in frequencies were not necessarily in the vicinity of the natural shedding frequencies, as previously observed in Figure~\ref{f:baselineres}. This disparity could be attributed to the large motion amplitudes, which significantly exceed the typical values observed for other shapes. Generally, maximum recorded amplitudes for e.g. rectangular [\cite{ZOU2025106031}] or circular cylinders [\cite{CHEN2020104119}] are smaller than their reference length. The high vertical velocities associated with the large displacement amplitudes of the present runs could drive a shift in the flow frequencies and a dislocation of the wake. This complicates the study of the phenomenon, preventing the prediction of critical VIV regions through a simple comparison of elastic and natural shedding frequencies.
The significant impact of the vertical displacement amplitude on the wake development is illustrated in Figure~\ref{f:flowviz}. The typical horizontally shed 2S pattern (Figure~\ref{f:flowviz1}) is replaced by an almost vertical development of the wake (Figure~\ref{f:flowviz2}), since the vertical velocity experienced by the airfoil is significant relative to the freestream velocity responsible for horizontal convection.

\subsection{Introduction of resonator}
\label{ch:resultsresonator}

\begin{figure*}[!htbp]
	\centering
	\includegraphics[width=\textwidth]{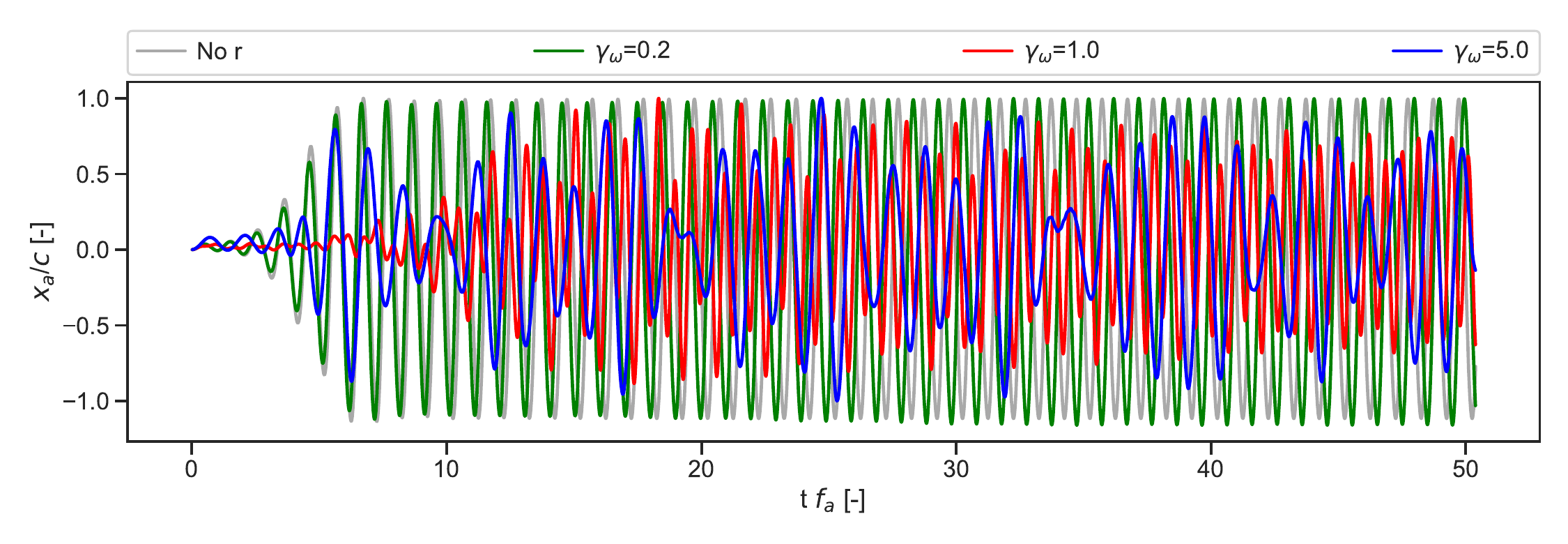}
	\caption{Effect of resonator on airfoil displacement for selected simulations at $U^*=5$ and $\gamma_m$=1.0. Time nondimensionalized with $f_a$ of the mounted configuration without resonator \texttt{No r}}
	\label{f:tseriesaddon}
\end{figure*}

Following the assessment of the VIV region for the baseline configuration (the elastically mounted airfoil), the introduction of a resonator was considered. Response studies focused on $U^*=5$, as this was the reduced velocity where the maximum amplitudes of motion were observed. To obtain an overview of the influence envelope on the vibrations, a parametric study was conducted. The three resonator parameters that were varied, each defined relative to the baseline system, were:
\begin{itemize}
	\item Mass ratio $\gamma_m = m_r / m_a$
	\item Stiffness ratio $\gamma_\omega = \omega_r / \omega_a$, with $\omega=\sqrt{k/m}$
	\item Damping ratio $\gamma_\xi = \xi_r / \xi_a$
\end{itemize}
In the study presented in this section $\gamma_\xi$ was fixed to 1.0, while the values of $\gamma_m$ and $\gamma_\omega$ were varied. A relatively broad range was selected for $\gamma_m$, including large (1.0), moderate (0.25) and low (0.05) values. Similarly, $\gamma_\omega$ was swept across a range from 0.2 to 100.0. To provide an overview of the influence of the resonator, Figure~\ref{f:tseriesaddon} shows the effect of the latter parameter for the specific configuration of $\gamma_m=1.0$. The effect of the resonator on the time series is manifested as a desynchronization of the characteristic frequencies of the problem, accompanied by a mitigation of the motion amplitudes.

\begin{figure}[!htbp]
	\centering
	\subfloat[Airfoil]{\includegraphics[width=0.5\textwidth]{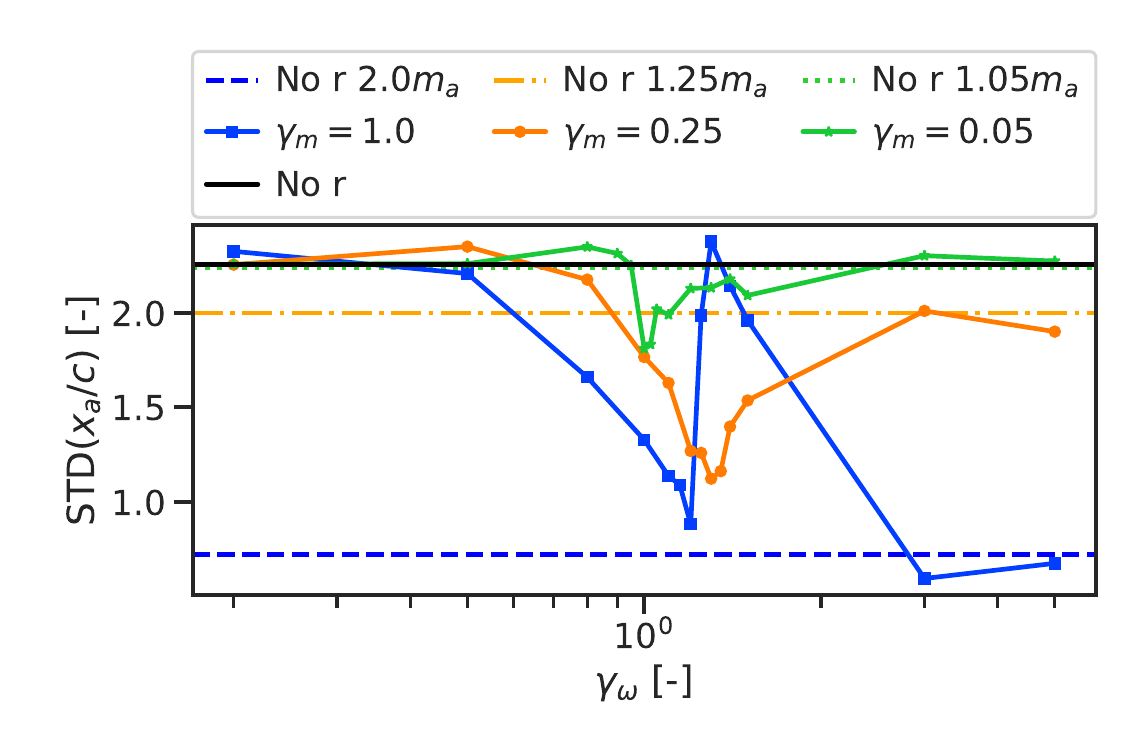}\label{f:dispoverviewaddow1}}
	\\
	\subfloat[Resonator]{\includegraphics[width=0.5\textwidth]{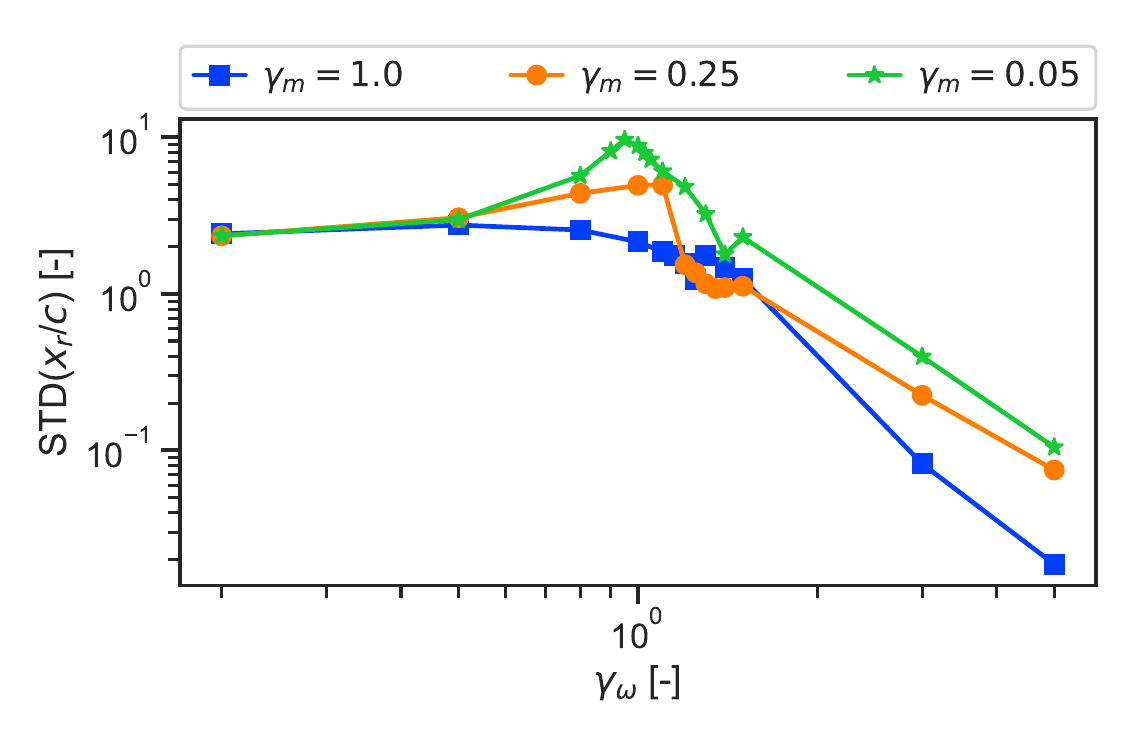}\label{f:dispoverviewaddow2}}
	\caption{Overview of STD of motion for selected cases with resonator at $\gamma_d=1.0.$}
	\label{f:dispoverviewaddow}
\end{figure}

\begin{figure}[!htbp]
	\centering
	\subfloat[Airfoil motion]{\includegraphics[width=0.33\textwidth]{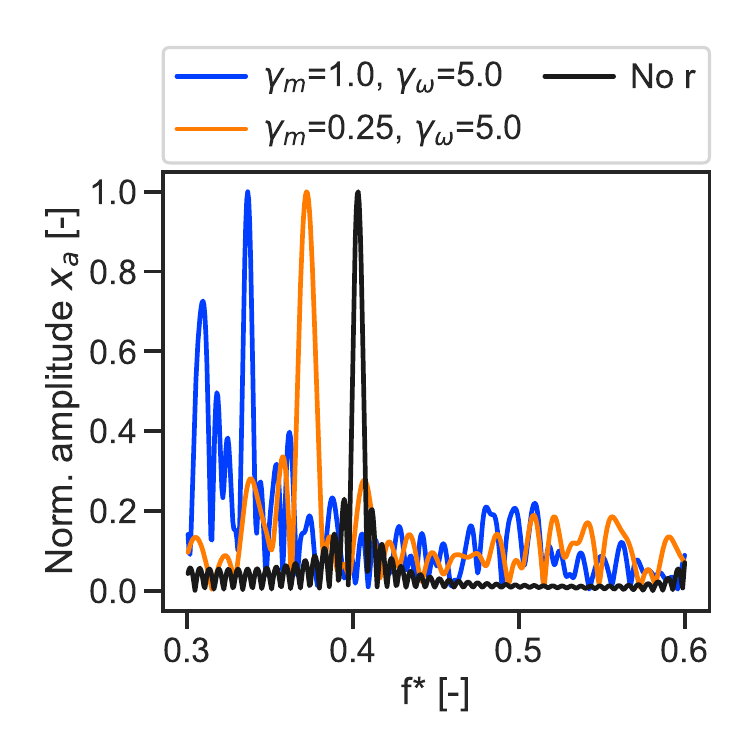}\label{f:highfreqfftanaly1}}
	\\
	\subfloat[Airfoil loads]{\includegraphics[width=0.33\textwidth]{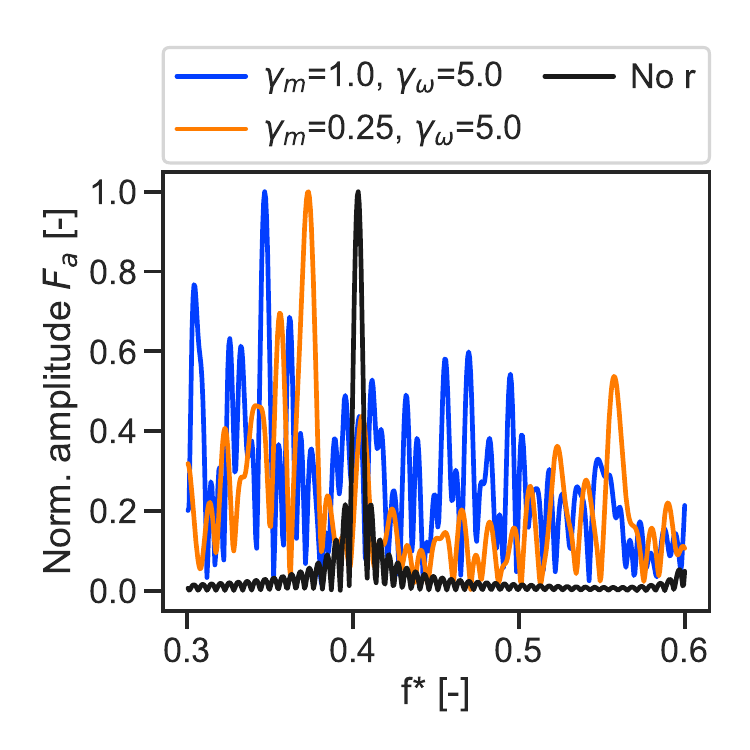}\label{f:highfreqfftanaly2}}
	\caption{Spectra for selected cases at $U^*=5$ and at high $\gamma_\omega$.}
	\label{f:highfreqfftanaly}
\end{figure}

A better characterization of these motion amplitudes is depicted in Figure~\ref{f:dispoverviewaddow}, where the results for selected values of $\gamma_m$ are presented as a function of $\gamma_\omega$. The standard deviation (STD) of the airfoil motion (Figure~\ref{f:dispoverviewaddow1}) demonstrates the ability of the resonator to attenuate a specific frequency band. For values of $\gamma_\omega$ below this band, the airfoil displacements correspond to those of a simulation without a resonator (marked as \texttt{No r} in the figure). Indeed, as $\omega_r$ becomes negligible, the loads exerted by the resonator (and therefore its capacity to influence the system) also become negligible.
On the other hand, the upper limit of the zone of influence of the resonator (i.e. high $\omega_r$ values) corresponds to configurations where the resonator is rigidly attached to the main system, as its relative period approaches zero (e.g. $\gamma_\omega$>3). Under these conditions, the resonator acts simply as an added mass on the main system. This behavior is also illustrated in Figure~\ref{f:dispoverviewaddow1}, by the addition of several simulations without a resonator but with a modified mass relative to the baseline airfoil mass $m_a$ (labeled as \texttt{No r} $\alpha m_a$). Consequently, when dealing with a non-negligible resonator mass, the displacement reduction achieved through its operational frequency band is superimposed onto this global attenuation driven by the added mass.
The latter effect is proportional to the resonator mass, as expected. Notably, for the specific case of $\gamma_m=1.0$, this mass increase alone is sufficient to bring the entire system out of lock-in. This is further illustrated in Figure~\ref{f:highfreqfftanaly}, which compares the originally synchronized motion of the configuration without a resonator (\texttt{No r}) with two resonator designs ($\gamma_m$=1.0 and $\gamma_m$=0.25) at very high $\gamma_\omega$. The spectra of airfoil displacements (Figure~\ref{f:highfreqfftanaly1}) show how the dominance of the frequency of motion is lost, especially for the heavier resonator. A better illustration of the lost of synchronization is found in the loads, as shown in Figure~\ref{f:highfreqfftanaly2}.
While both the added mass and the frequency band influence of the resonator successfully reduce vibration amplitudes, a clear drawback is the exceptionally large, potentially unphysical displacement of the resonator itself (Figure~\ref{f:dispoverviewaddow2}).
Although adjusting the damping value (which was fixed to $\gamma_\xi=1.0$ here) could successfully reduce the oscillation amplitude of the resonator, it did so at the expense of the overall efficiency in mitigating the vibrations of the airfoil. A dedicated analysis of this damping trade-off is provided as a complementary study in Appendix~\ref{ch:damping}.

\begin{figure}[!htbp]
	\centering
	\subfloat[$\gamma_\omega$=0.8]{\includegraphics[width=0.33\textwidth]{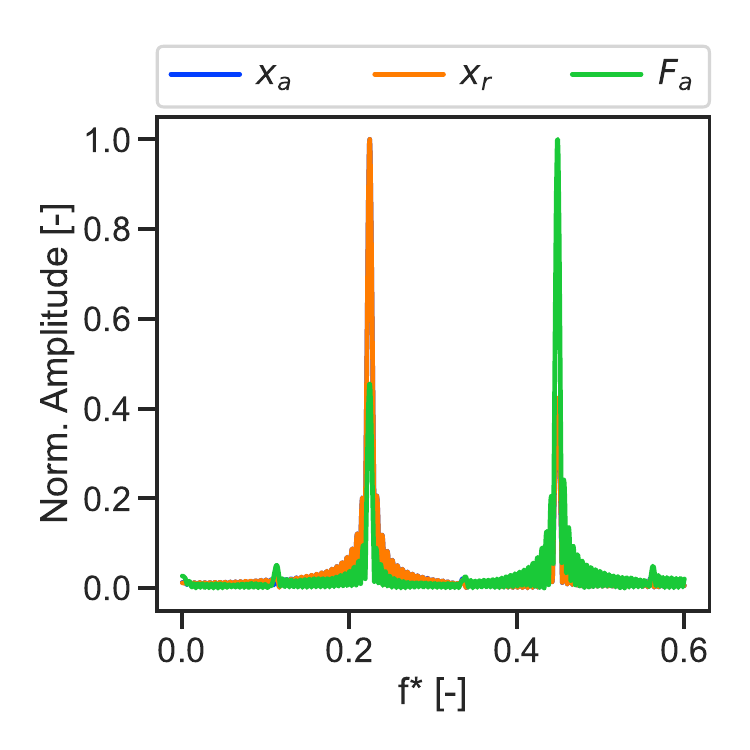}\label{f:medfreqaddonfreqs1}}
	\\
	\subfloat[$\gamma_\omega$=1.2]{\includegraphics[width=0.33\textwidth]{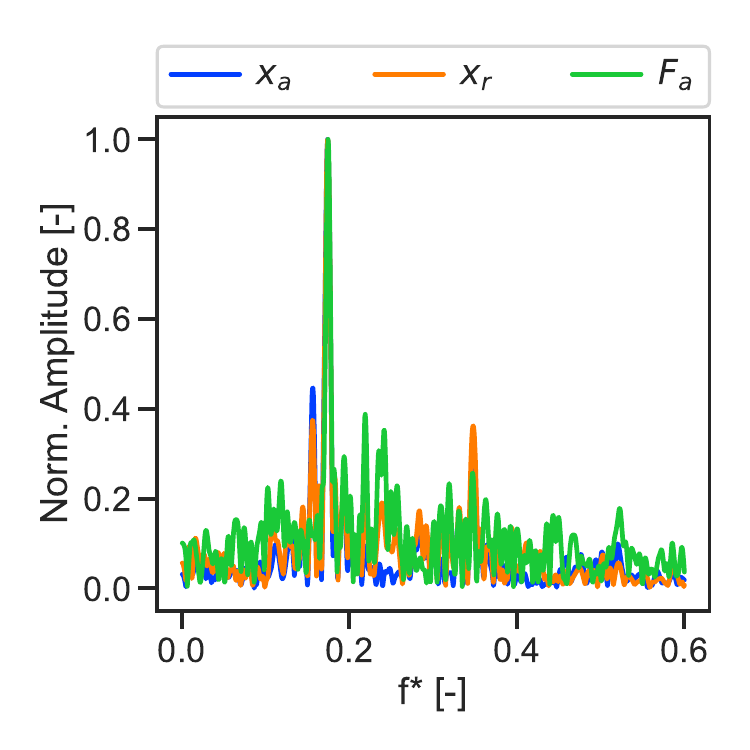}\label{f:medfreqaddonfreqs2}}
	\caption{Normalized loads and displacements spectra for selected cases in the resonator area of incluence. All cases assume $\gamma_m$=0.25 and consider $U^*=5$.}
	\label{f:medfreqaddonfreqs}
\end{figure}

The minimum vibration amplitudes are summarized in Table~\ref{t:maxvibrations}, which also includes the relative mitigation with respect to the baseline simulation without a resonator (\texttt{No r}). The results indicate that while an amplitude attenuation of more than 50\% is achievable, this is only possible when employing large mass ratios ($\gamma_m > 0.25$).
\begin{table}[!htbp]
	\centering
	\caption{Minimum $x_a$ oscillations in resonator frequency band. Assumed $\gamma_\xi=1$. Mitigation refers to the relative percentual value with respect to the \texttt{No r} configuration.}
	\label{t:maxvibrations}
	\begin{tabular}{llll}
		\toprule
		                                          & $\gamma_m=0.05$ & $\gamma_m=0.25$ & $\gamma_m=1.0$ \\
		\midrule
		$\text{STD} \left ( x_a \right ) / c$ [-] & 1.81            & 1.12            & 0.88           \\
		Mitigation [\%]                           & 19.6            & 50.2            & 60.9           \\
		\bottomrule
	\end{tabular}
\end{table}
While this highlights that the mitigation capabilities of the resonator are limited by its mass (at least in the present configuration), it is worth studying the frequency mechanisms activated within the frequency band of influence of the resonator to further exploit them in the future. It is within this region that the resonator and the airfoil exchange energy at the same aeroelastic frequency. This frequency was found to shift significantly when varying $\gamma_\omega$, as illustrated in Figure~\ref{f:medfreqaddonfreqs}. The results highlight the effectiveness of the resonator, demonstrating its ability to bring the originally locked-in system (Figure~\ref{f:medfreqaddonfreqs1}) completely out of that state (Figure~\ref{f:medfreqaddonfreqs2}).

\begin{figure}[!htbp]
	\centering
	\subfloat[$\gamma_m$=0.05]{\includegraphics[width=0.33\textwidth]{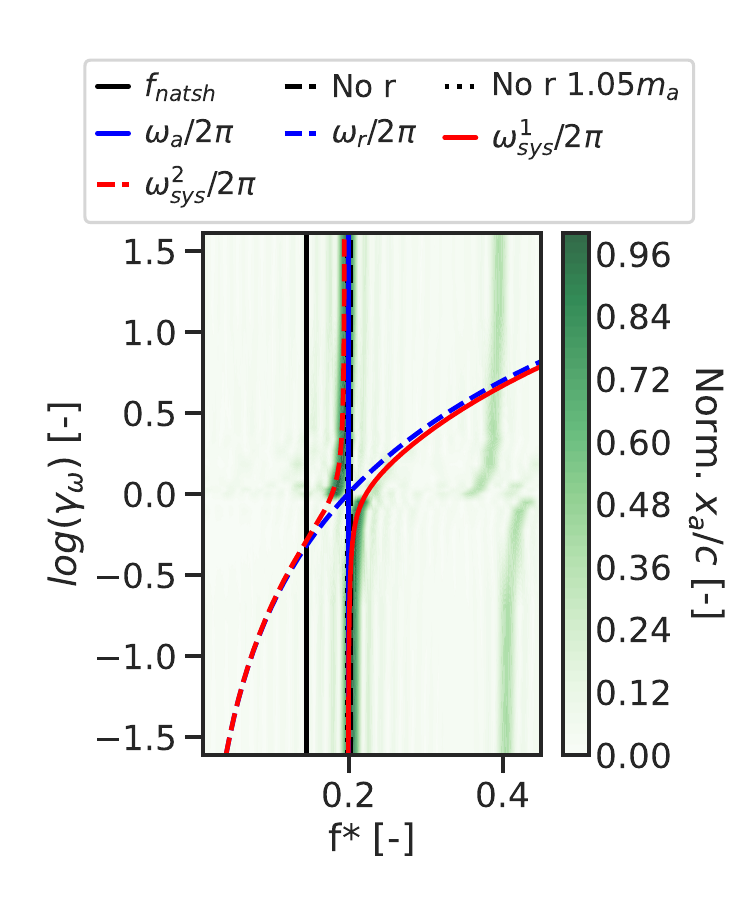}\label{f:characfreq11}}
	\\
	\subfloat[$\gamma_m$=0.25]{\includegraphics[width=0.33\textwidth]{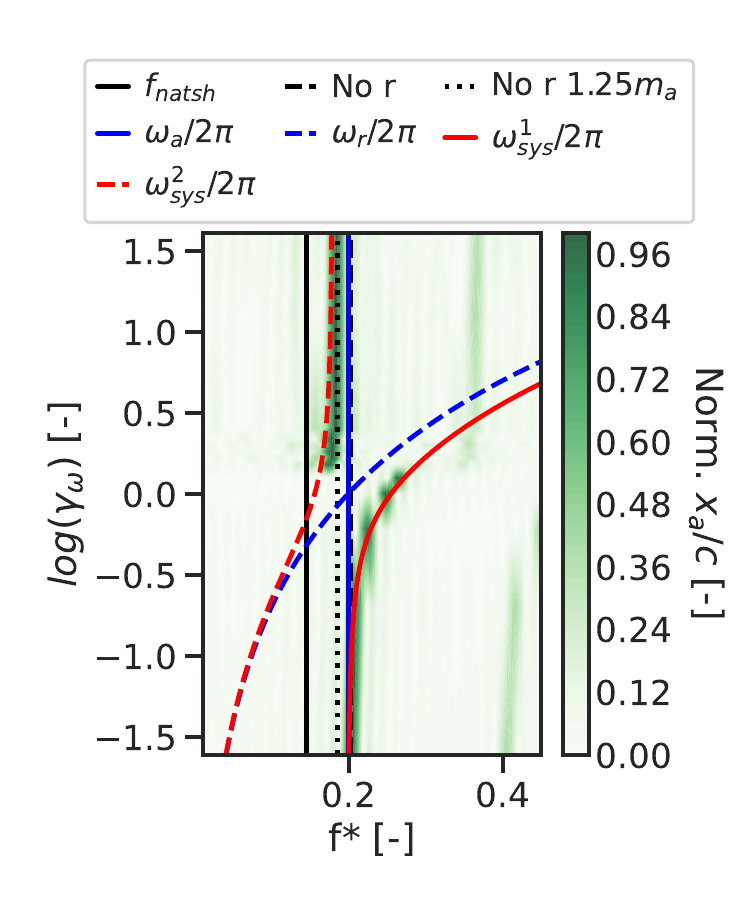}\label{f:characfreq12}}
	\\
	\subfloat[$\gamma_m$=1.0]{\includegraphics[width=0.33\textwidth]{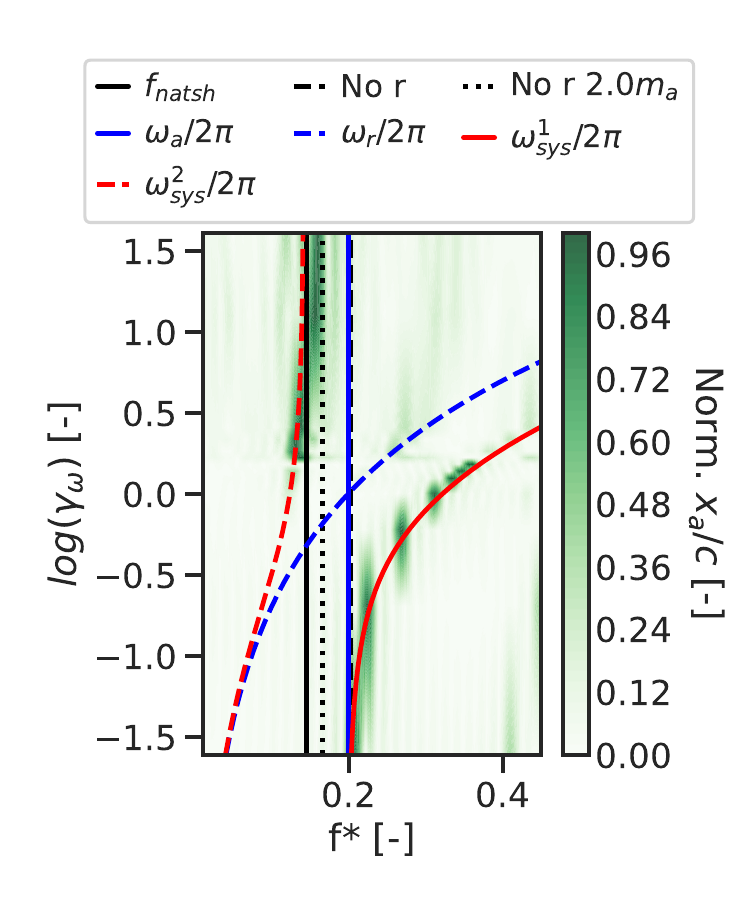}\label{f:characfreq13}}
	\caption{Spectra of $x_a$ for $U^*=5$, assuming $\omega_\xi=1$. Results normalized by the maximum of every row.}
	\label{f:characfreq}
\end{figure}

A better characterization of the frequency shifting mechanism, which is assumed to drive the mitigation capabilities of the resonators, is presented in Figure~\ref{f:characfreq}. The figure stacks the displacement spectra of $x_a$ for every studied $\gamma_\omega$ at a given $\gamma_m$. The uncoupled structural frequencies of the elastic mounting $\omega_a$ and of the resonator $\omega_r$ are included for reference. As it can be seen, the problem is better described by the coupled structured frequencies (i.e. the frequency of the \texttt{Airfoil} plus resonator system). Such variables, labeled as $\omega^1_{sys}$ and $\omega^2_{sys}$, were found by solving for the natural frequencies of the coupled undamped and unloaded system, $m_a m_r \omega^4 - \left[m_a k_r + m_r (k_a + k_r)\right] \omega^2 + k_a k_r = 0$.
The figure reveals how the spectra of the airfoil displacements follow the split of these coupled structural frequencies, as the values of $f^*$ with high $x_a$ amplitudes track this separation. Additionally, it can be observed that the resonator frequency band of influence corresponded to regions of large separation between the coupled structural frequencies and the characteristic frequencies of the problem without a resonator (i.e. natural shedding frequency $f_{shnat}$ and the oscillation frequencies without a resonator, \texttt{No r}). An analysis of the analogous spectra for $C_l$, not included here for brevity, also revealed the disappearance of lock-in for cases where the amplitude attenuation was most pronounced.
Therefore, the intensity of the mitigation seems to depend directly on the ability of the system to establish a wide gap between the various frequencies involved. This explains why the amplitude mitigation capabilities of the heavier resonators (Figure~\ref{f:characfreq11} and Figure~\ref{f:characfreq12}) were much more significant than in the case of $\gamma_m=0.05$ (Figure~\ref{f:characfreq13}).

\begin{figure}[!htbp]
	\centering
	\includegraphics[width=0.5\textwidth]{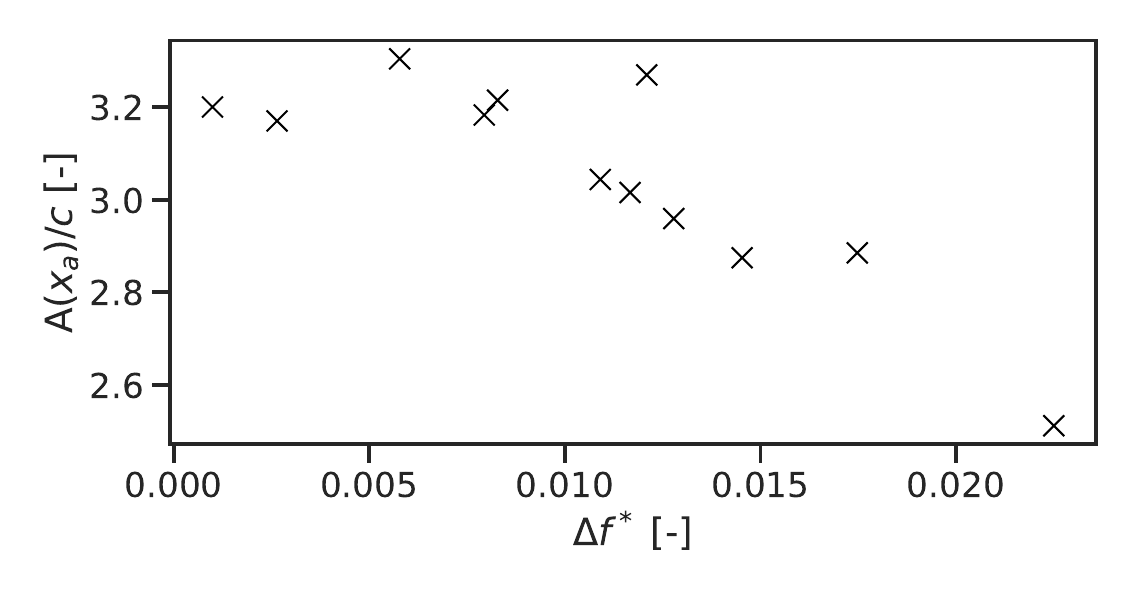}
	\caption{Airfoil modal amplitude against induced frequency departure from mounting frequency $\Delta f^*$. Using $\gamma_m$=0.05 and $U^*=5$}
	\label{f:deltafandA}
\end{figure}

The relationship between the ability of the system to mitigate vibration amplitudes and its capacity to alter the underlying frequencies is depicted in Figure~\ref{f:deltafandA} for the specific case of $\gamma_m=0.05$. The modal amplitude around the main motion frequency of the airfoil is plotted as a function of $\Delta_f^*$. This metric serves as an indicator of the ability of the resonator to bifurcate the structural frequencies. Specifically, $\Delta_f^* = \frac{1}{2 \pi} \min \left( \left| \omega_{\mathrm{sys}}^{1} - \omega_a \right|, \left| \omega_{\mathrm{sys}}^{2} - \omega_a \right| \right)$. The results reveal a clear dependency between amplitude reduction and frequency shifting, despite an observed asymmetry between the branches before and after the minimum.

\section{Conclusions and future work}

A computational methodology was developed for a preliminary assessment of the potential influence of locally resonant metamaterials on VIV. The scope of the problem was directed toward large industrial structures, specifically wind turbine blades. Although the proposed setup represented a considerable simplification, it successfully captured the nonlinearity of the problem, revealing both the underlying physical mechanisms and their limitations. Specifically, a 2D representation was adopted, reducing the complex 3D scenario to the classical problem of VIV on an airfoil undergoing transverse motion. The modeling of the resonators was also simplified by representing them as an additional spring and a damper, coupled to the main system. A state-of-the-art finite volume code was used to model the flow and solve the FSI problem. The capabilities of this solver were extended to account for the resonators through a tailored modification based on finite differences.

Following the validation of the methodology against reference VIV experiments, the influence of the design parameters of the resonator was analyzed. Particular attention was given to the mass of the resonator relative to the airfoil, as well as the characteristic frequencies of the system. The study identified a specific frequency region of influence for the resonator (i.e. a frequency band), where efficient vibration suppression was achieved. The resonator was found to drive the system out of lock-in through the bifurcation of the structural frequencies induced by its addition. As such, this mitigation mechanism was intrinsically limited by the mass of the resonator relative to the main airfoil. Specifically, achieving an attenuation of more than 50\% of the original oscillation required a relative resonator mass of at least 25\% compared to the baseline case.

The uncovered mechanisms pave the way for the exploitation of frequency bifurcation for vibration mitigation in future studies. These future efforts will initially center on the most limiting factor identified: the required resonator mass. In this regard, it should be noted that the selected VIV experiments exhibited significantly high vibration amplitudes. Such oscillations are larger than what is traditionally observed in wind energy applications, or even for most of the canonical circular cylinder analyses (typically limiting at 0.6 diameters, as for \cite{LUO2022103692}). Therefore, the efficiency of adding a local resonator must be reassessed under more representative circumstances.
Alternatively, studies based on a total mass preserving approach (where mass is transferred from the main system to the resonator rather than being added) could further maximize the efficiency of the proposed device. Finally, to increase the fidelity of the model, future work could transition from modeling the metamaterial as a single resonator to representing it as a series of distributed resonators, as relying on a single macroscopic resonator may inherently limit the operational frequency band over which the device is efficient.

\appendix

\section{Verification of aeroelastic coupling}
\label{ch:verification}

\begin{figure}[!htbp]
	\centering
	\subfloat[Only airfoil (baseline \texttt{OpenFOAM})]{\includegraphics[width=0.5\textwidth]{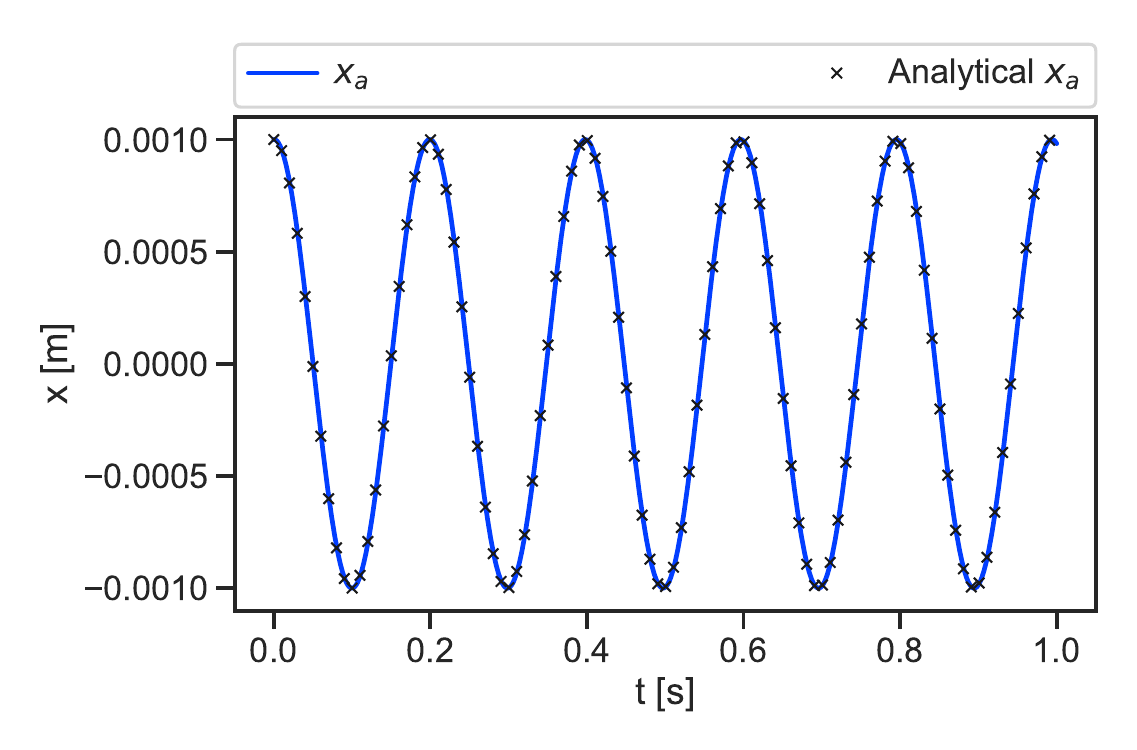}\label{f:verification1}}
	\\
	\subfloat[Airfoil and resonator implementation]{\includegraphics[width=0.5\textwidth]{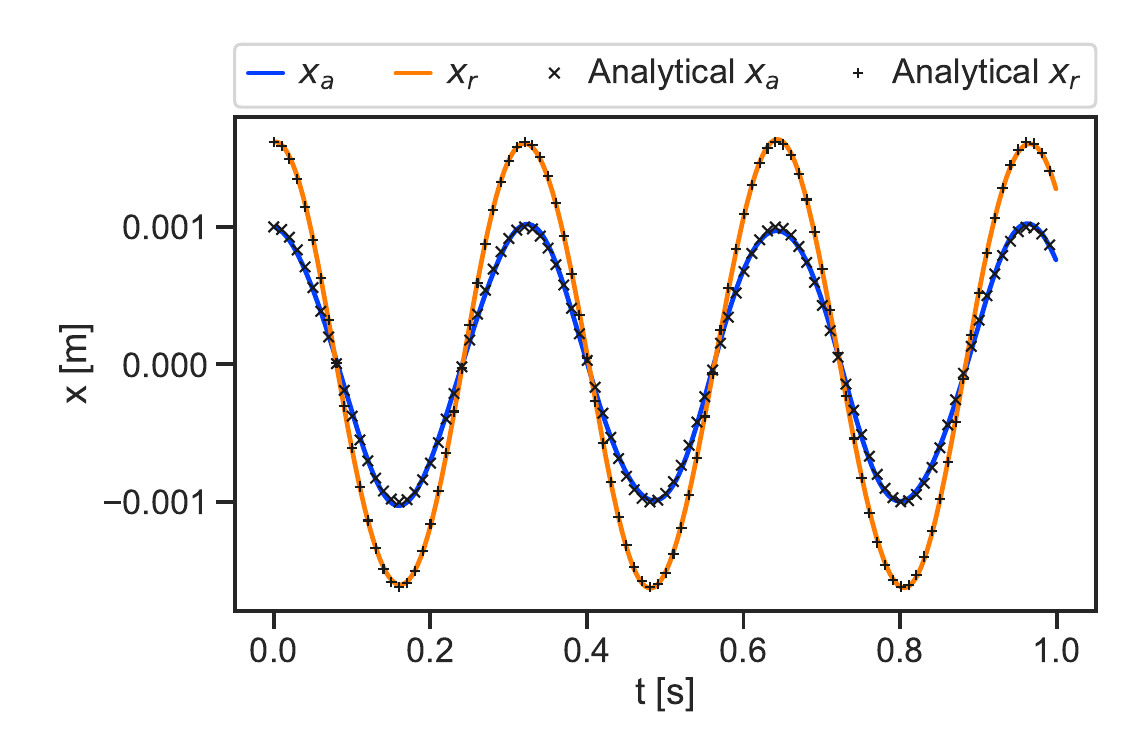}\label{f:verification2}}
	\caption{Verification of airfoil mounting and resonator.}
	\label{f:verification}
\end{figure}

The proposed implementation of the elastic mounting was verified against analytical solutions derived for specific configurations (Figure~\ref{f:verification}). These solutions were obtained by assuming zero external aerodynamic forcing ($F_a=0$) and neglecting damping. For simplicity, the initial conditions were defined such that only a single mode shape was excited. Both the default capabilities of \texttt{OpenFOAM} (i.e. in the absence of a resonator) and the newly implemented coupled system were verified, as shown in Figure~\ref{f:verification1} and Figure~\ref{f:verification2}, respectively. In both cases, the main system parameters were set to $k_a=10^3$ N/m and $m_a=1$ kg. For the configuration that included the resonator, the conditions $k_r=k_a$ and $m_r=m_a$ were imposed to allow for a straightforward computation of the analytical solution. The comparison of the two figures illustrates the shift in the characteristic frequencies of the system induced by the introduction of the resonator.

\section{Validation of unmounted configuration}
\label{ch:valnomounted}

\begin{figure}[!htbp]
	\centering
	\subfloat[Lift coefficient for FFA w3 241 airfoil]{\includegraphics[width=0.33\textwidth]{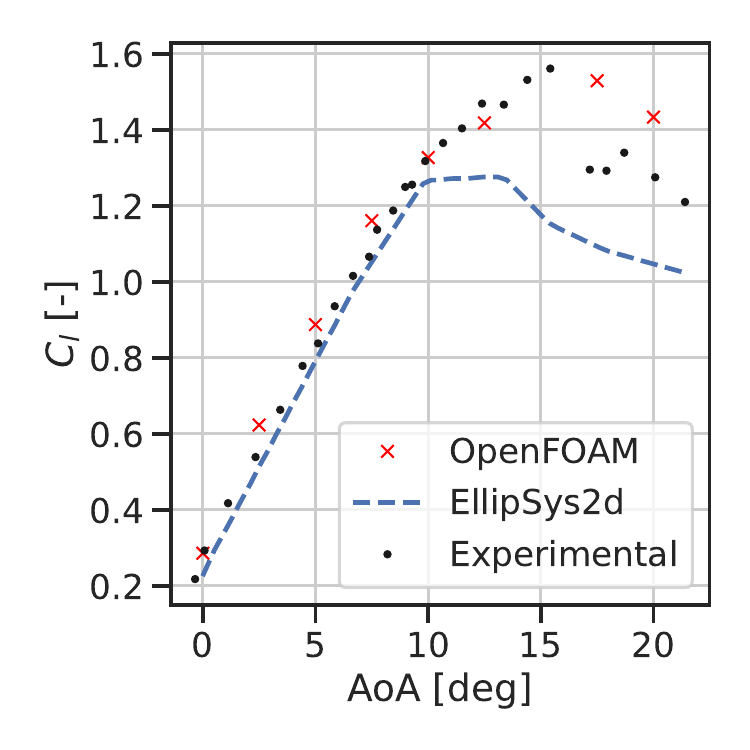}\label{f:validations1}}
	\\
	\subfloat[Drag coefficient for FFA 241 airfoil]{\includegraphics[width=0.33\textwidth]{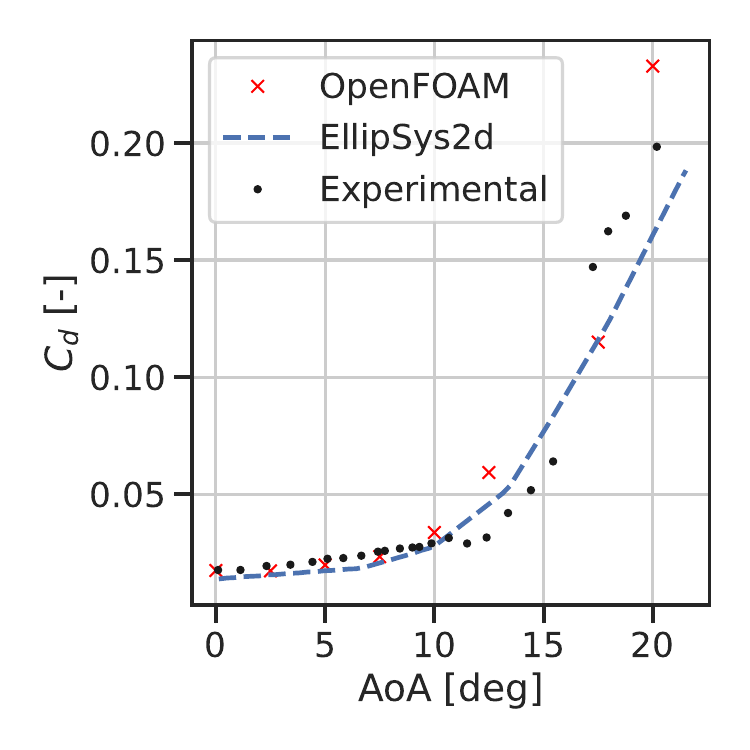}\label{f:validations2}}
	\\
	\subfloat[Pressure distribution for NACA0012]{\includegraphics[width=0.33\textwidth]{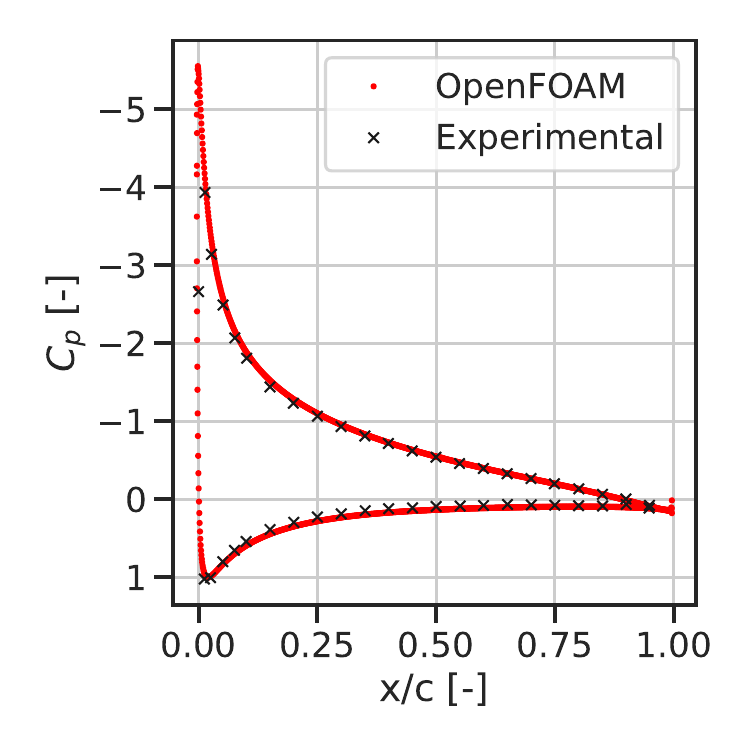}\label{f:validations3}}
	\caption{Selected validations for unmounted configuration.}
	\label{f:validations}
\end{figure}

Prior to the study of the airfoil at \ang{90} and the analysis of the elastic mounting, the numerical setup was assessed against literature data using the steady RANS variant of the model, accounting for both attached and separated flow conditions (Figure~\ref{f:validations}). Comparisons were made against other similar software assuming a fully turbulent boundary layer, as well as experimental results under tripped conditions. These reference studies followed a mesh refinement strategy similar to the present work, with the exception of the boundary layer mesh, which was specifically tailored to the targeted Reynolds number.
Overall, the code demonstrated good agreement with experimental data regarding the integrated loads of a typical wind turbine airfoil. This is illustrated by the lift and drag coefficients in Figure~\ref{f:validations1} and Figure~\ref{f:validations2}, respectively, for the FFA-W3-241 airfoil at a Reynolds number of 1.5 million (\cite{wtcatalogue}). The results reveal a good match not only in the linear region of the airfoil but also provide a relatively accurate prediction of the maximum lift coefficient and stall angle. Results from the finite volume code EllipSys2D (which is widely used and validated for airfoil simulations) are included, demonstrating consistency with the \texttt{OpenFOAM} results obtained in this study.
This strong agreement was observed not only for integrated quantities but also for distributed metrics. As illustrated in Figure~\ref{f:validations3}, the pressure coefficient distribution around a NACA0012 airfoil (\cite{Ladson1987HighRN}) predicted by \texttt{OpenFOAM} shows an excellent fit with the experimental data at an angle of attack of \ang{10} and a Reynolds number of 9 million.

\section{Sensitivities of aeroelastic model}
\label{ch:sensitivities}

\begin{figure}[!htbp]
	\centering
	\subfloat[Mesh refinement (with baseline max. Courant)]{\includegraphics[width=0.45\textwidth]{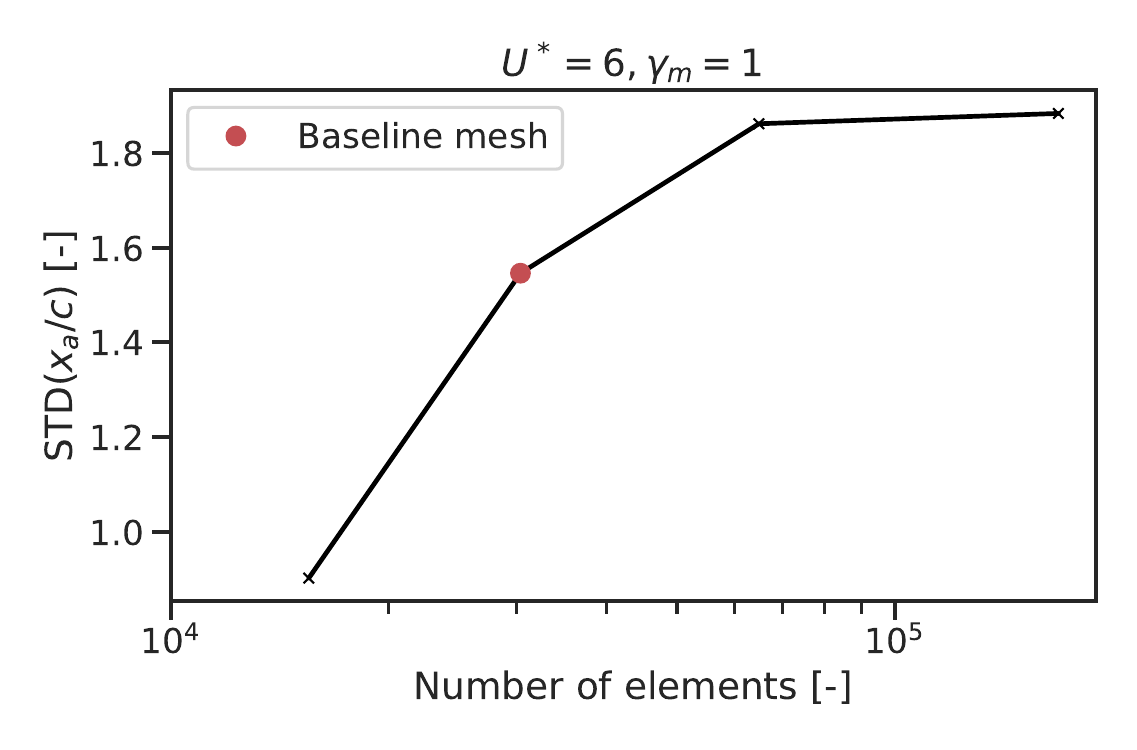}\label{f:sensi1}}
	\\
	\subfloat[Time step length (with baseline mesh)]{\includegraphics[width=0.45\textwidth]{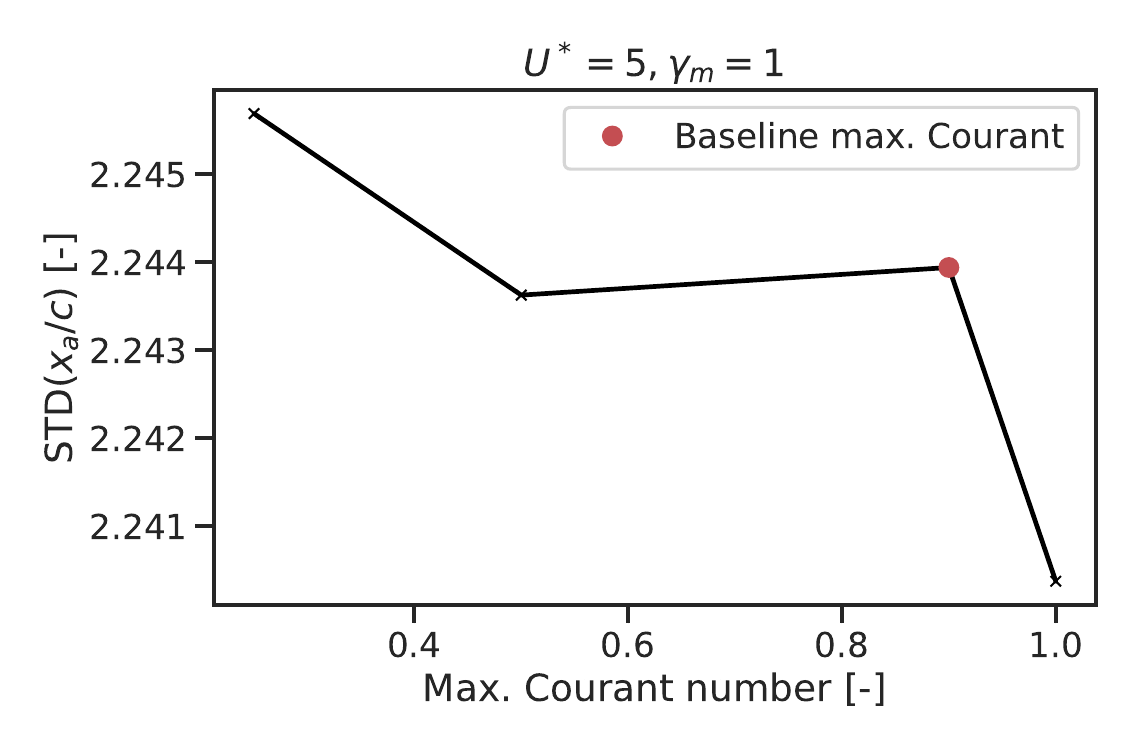}\label{f:sensi2}}
	\caption{Selected sensitivity studies. Standard deviation of airfoil displacement (without resonator).}
	\label{f:sensi}
\end{figure}

Prior to validating the aeroelastic model and exploring the effects of the resonator, several sensitivity studies were performed for the baseline VIV problem. The most critical of these (grid resolution and time step) are included in Figure~\ref{f:sensi} for reference.
Specifically, Figure~\ref{f:sensi1} depicts a typical mesh sensitivity analysis based on the displacements experienced by the airfoil. This was conducted by systematically reducing the characteristic cell size of the CFD domain (with the exception of the boundary layer region, that was kept fixed to ensure a consistent comparison). Although the model demonstrated good convergence with further refinement, the chosen mesh resolution was set slightly below this fully stabilized region, as a trade-off for computational efficiency. This compromise aimed at better exploiting the resources, and allowed for a broader exploration of the parametric space.
In contrast, the model exhibited significantly lower sensitivity to the chosen time step. This robustness is illustrated in Figure~\ref{f:sensi2} through variations in the maximum Courant number.

\section{Influence of damping on resonators mitigation efficiency}
\label{ch:damping}

\begin{figure}[!htbp]
	\centering
	\subfloat[Airfoil]{\includegraphics[width=0.5\textwidth]{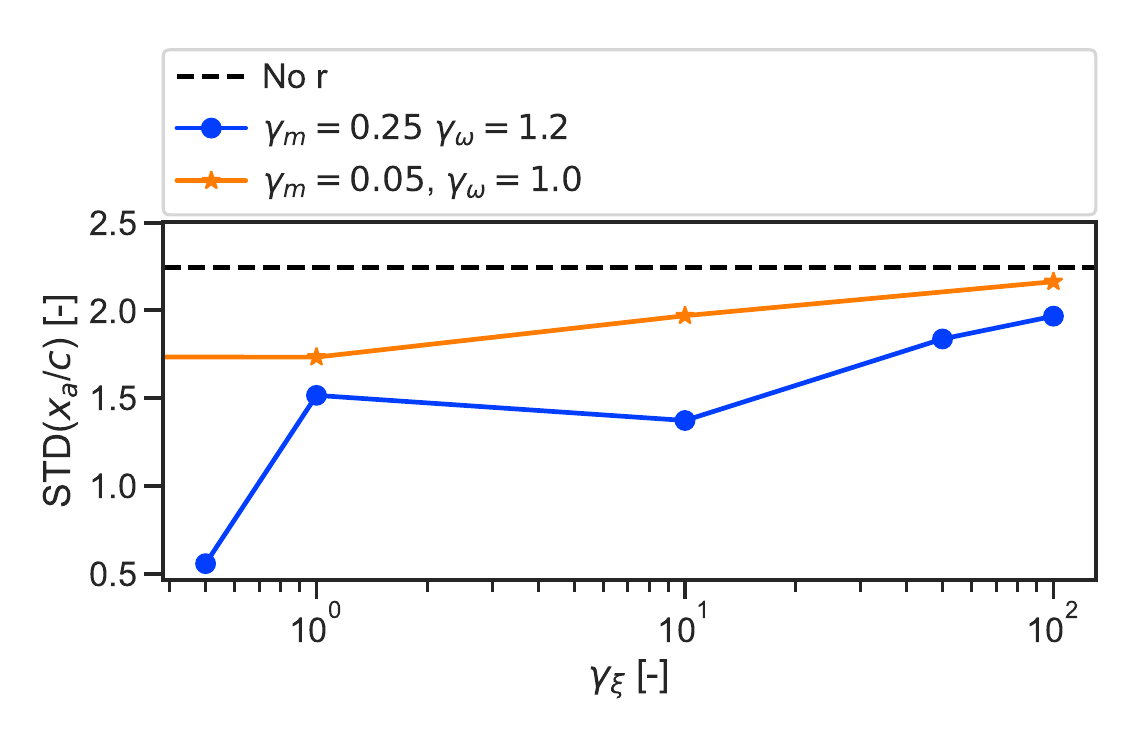}\label{f:medfreqaddondamping1}}
	\\
	\subfloat[Resonator]{\includegraphics[width=0.5\textwidth]{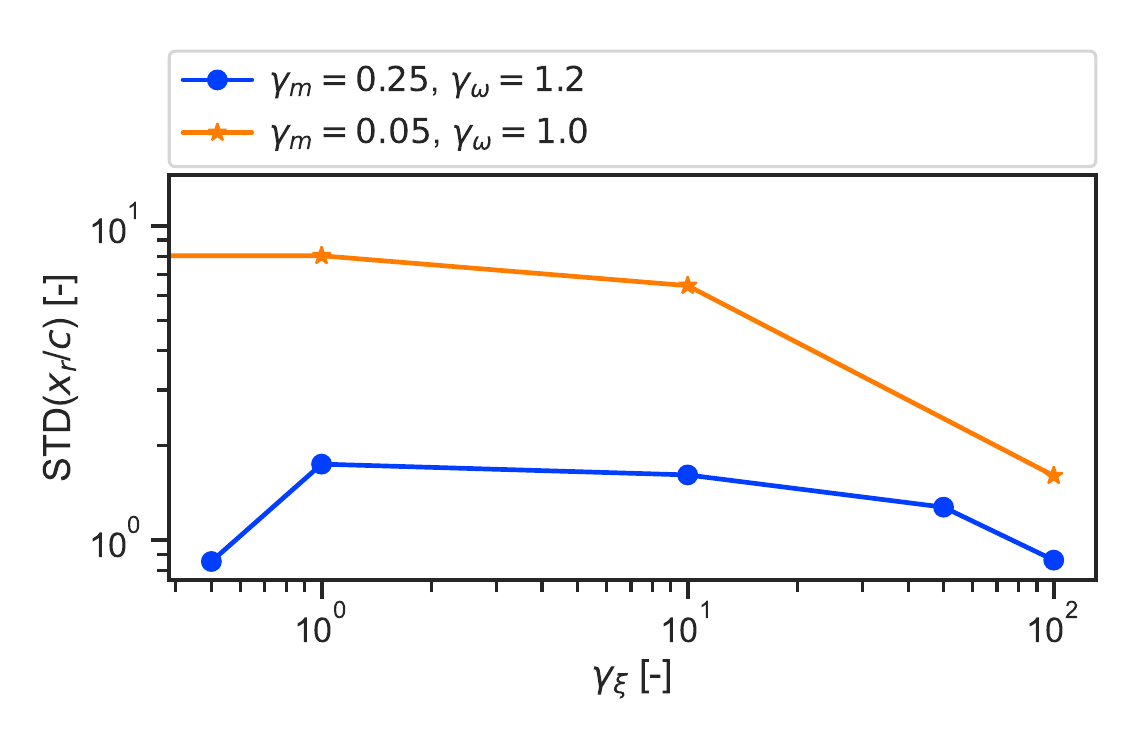}\label{f:medfreqaddondamping2}}
	\caption{Effect of damping on amplitude of motion. Selected configurations}
	\label{f:medfreqaddondamping}
\end{figure}

A parametric study on the relative damping coefficient $\gamma_\xi$ was conducted as an independent analysis to shed light on its influence on the response of the system. Specifically, $\gamma_\xi$ was varied for two configurations that led to maximum attenuation: ($\gamma_m=0.25, \gamma_\omega=1.2$) and ($\gamma_m=0.05, \gamma_\omega=1.0$).
The results of this study, summarized in Figure~\ref{f:medfreqaddondamping}, demonstrate that the value of $\gamma_\xi$ governs the ratio between the displacement amplitude of the resonator and that of the airfoil. In particular, higher values of $\gamma_\xi$ reduce the vibration mitigation efficiency of the airfoil (Figure~\ref{f:medfreqaddondamping1}), but they simultaneously restrict the oscillation amplitude of the resonator itself (Figure~\ref{f:medfreqaddondamping2}).
This finding highlights the importance of exploring alternative configurations aimed at reducing resonator stroke, such as recent studies involving the introduction of an inerter [\cite{ZHAO2026126216}].

\section*{Acknowledgments}
During the preparation of this manuscript, the authors used Google Gemini to improve the grammar and language of the text. The authors reviewed and edited the output as necessary and take full responsibility for the content of the published article.

S.G. Horcas has received funding from the postdoctoral fellowship programme Beatriu de Pinós, funded by the Secretary of Universities and Research (Government of Catalonia), under grant 2022-BP-00260.
D. Roca is a Serra Hunter fellow.
The authors would also like to acknowledge the Turbulence and Aerodynamics in Mechanical and Aerospace Engineering Research Group (TUAREG) at Universitat Politècnica de Catalunya, particularly A. Miro and M. Soria, for providing the necessary computational resources to perform this study.

\section*{AUTHOR DECLARATIONS}

\subsection*{Conflict of Interest}

The authors have no conflicts to disclose.

\subsection*{Author Contributions}
S. G. Horcas:
Project administration, Funding acquisition, Conceptualization,
Methodology, Software, Validation, Investigation, Formal analysis,
Writing - original draft.
D. Roca:
Conceptualization, Methodology, Software, Formal analysis,
Writing - original draft.
E. Ortega:
Methodology, Supervision, Writing - review \& editing.
J. Cante:
Resources, Methodology, Conceptualization,
Writing - review \& editing.

\section*{DATA AVAILABILITY}

The data that support the findings of this study are available
from the corresponding author upon reasonable request.

\bibliographystyle{plainnat}
\bibliography{library}

\end{document}